# Nonlinear investigation of the pulsational properties of RR Lyrae variables

G. Bono,[1] F. Caputo,[2] V. Castellani[3,4,5] and M. Marconi[3]

[1] Osservatorio Astronomico di Trieste, Via G.B.Tiepolo 11, 34131 Trieste, Italy; E-mail: bono@oat.ts.astro.it
[2] Istituto di Astrofisica Spaziale, CNR, C.P. 67, 00044 Frascati, Italy; E-mail: caputo@saturn.ias.fra.cnr.it
[3] Dipartimento di Fisica, Università di Pisa, Piazza Torricelli 2, 56126 Pisa, Italy; E-mail: marcella@astr1pi.difi.unipi.it
[4] Osservatorio Astronomico Collurania, 64100 Teramo, Italy; E-mail: vittorio@astr1pi.difi.unipi.it
[5] Laboratori del Gran Sasso, INFN, 67100 L'Aquila, Italy

June 5, 1996

**Abstract.** We present a theoretical investigation on periods and amplitudes of RR Lyrae pulsators by adopting stellar parameters which cover the range of theoretical evolutionary expectations. Extensive grids of nonlinear, nonlocal and time-dependent convective RR Lyrae envelope models have been computed to investigate the pulsational behavior in both fundamental and first overtone modes at selected luminosity levels and over an effective temperature range which covers the whole instability region. In order to avoid spurious evaluations of modal stability and pulsation amplitudes, the coupling between pulsation and convection was followed through a direct time integration of the leading equations until radial motions approached their limiting amplitude.
Blue and red boundaries for pulsational instability into the HR diagram are presented for three different mass values M=0.75, 0.65 and 0.58 $M_\odot$, together with an atlas of full amplitude theoretical light curves for both fundamental and first overtone pulsators and for two different assumptions of stellar masses: M=0.75 and 0.65 $M_\odot$ [1]. The dependence of periods on stellar parameters is discussed and new analytical relations connecting the period to the masses, luminosities and effective temperatures are provided.
We show that theoretical expectations concerning minimum fundamental periods are in good agreement with the observational evidence of a dichotomic period distribution between different Oosterhoff type clusters. A rather good correlation has been found between the pulsational amplitude of fundamental pulsators and the effective temperature, rather independently of stellar mass and luminosity. Theoretical periods have been combined with theoretical amplitudes in order to predict the location of the pulsators in the Bailey amplitude-period diagram. Comparison with observational data brings to light what we regard as a clear indication that the OR region, *i. e.* the region where both fundamental and first overtone show a stable limit cycle, is populated by fundamental or first overtone pulsators in Oosterhoff I and Oosterhoff II clusters respectively. Some evident mismatches between theory and observation have also been found, and they are presented and discussed.

## 1. Introduction

The pulsational properties of RR Lyrae variables of galactic globular clusters (GGCs) provide an unique opportunity for testing the structural parameters of pulsating stars and the prescriptions of stellar evolutionary theories. According to such an evidence, in the last decades the well known van Albada & Baker (1971) relation connecting the pulsation periods to stellar masses, luminosities and effective temperatures has been at the basis of several investigations on the evolutionary status of HB stars, raising in particular a well known debate about the actual luminosity of RR Lyrae pulsators.

A relevant point to remember is that pulsational periods are unaffected either by interstellar reddening or by the distance modulus and thus provide firm and accurate observational data which unambiguously constrain the evolutionary parameters of the evolving variables. However, the pulsational phenomenon provides a further relevant observable, namely the amplitude of the luminosity variation. According to such plain evidence, as early as the beginning of the century Bailey (1899, 1902) investigated the properties of RR Lyrae pulsators in the period-amplitude diagram which is now usually named after its inventor.

---

However, the investigation of pulsational properties in the Bailey diagram has never gone beyond an empirical and qualitative approach. For instance, we know that along this diagram the RR Lyrae variables of different Oosterhoff types show different distributions, but until recently we have been unable to investigate the origin of such a behavior because we lack a detailed theoretical scenario connecting pulsational amplitudes with the structural parameters of the pulsating stars. This is due to the fact that periods can be easily derived by a simple linear and adiabatic approach to the pulsational phenomenon, i. e. the canonical von Ritter relation, whereas the amplitudes are the result of detailed nonlinear investigations which require a much more difficult theoretical and computational effort.

To address this problem properly, during the last few years we have undertaken a theoretical project aimed at the evaluation of the pulsational amplitudes and the modal stability of RR Lyrae variables by computing a large and homogeneous set of nonlinear and time-dependent convective envelope models. The primary goal of this approach is to provide, for the first time, a sound theoretical scenario of the dependence of all the pulsational observables on astrophysical parameters such as stellar mass and chemical composition. Moreover, the pulsation properties of nonlinear RR Lyrae models, when evaluated at full amplitude, can supply useful suggestions concerning the physical processes which drive and/or quench the pulsation mechanism.

Bono & Stellingwerf (1994, hereinafter referred to as BS) have already shown that a nonlinear approach is necessary for an accurate determination of periods. The outcome was similar for properly locating the instability regions into the HR diagram. Indeed, the nonlinear instability boundaries of fundamental (F) and first overtone (FO) pulsators have been already found in substantial agreement with the observed color distribution of RR Lyrae variables in selected galactic globulars (Bono, Caputo & Marconi 1995, hereinafter referred to as BCM). Moreover, the minimum fundamentalized periods of RR pulsators belonging both to Oosterhoff I (OoI) and Oosterhoff II (OoII) globular clusters have been explained by taking simultaneously into account the evolutionary history and the modal stability of variable stars in the instability strip (Bono et al. 1995).

The results presented in this paper have been obtained by using the same theoretical framework adopted by Bono & Stellingwerf (1992) and BS in the first two papers of this series. Physical assumptions and numerical procedures have been already described in the quoted papers and in Stellingwerf (1975, 1982). The reader is referred to these papers for a detailed description. The main differences between the present investigation and the results obtained by BS are:
1) the RR Lyrae models supplied by BS were computed by assuming a fixed mass value (M=0.65 $M_\odot$) whereas in the present paper three different mass values have been adopted (M=0.58, 0.65, 0.75 $M_\odot$) which cover the whole expected range of HB stars.
2) In order to compare the pulsation properties of RR Lyrae models with previous theoretical works, BS adopted a helium abundance by mass of Y=0.30 whereas all the present computations have been performed by adopting a helium content of Y=0.24. We chose this value since on the basis of both the R (or R') method (Buzzoni et al. 1983) and the determination of the ratio He/H in HII regions and planetary nebulae (Peimbert & Torres-Peimbert 1976; Peimbert 1995 and references therein) there is a general consensus on the fact that a suitable range for the helium abundance of population II stars belonging to GGCs is $0.22 \leq Y \leq 0.25$ (Pagel 1995 and references therein).

Owing to the marginal dependence of the pulsational properties of metal poor stars on metal content (Christy 1966), a fixed metal abundance, Z=0.0001, has been adopted. However, it is worth underlining that throughout the pulsation cycle the transformation of luminosity and temperature into the observational plane depends on the metal content, since both bolometric correction and color-temperature relation present a non negligible dependence on such a parameter.

According to theoretical expectations (see, e.g., Castellani, Chieffi & Pulone 1989, hereinafter referred to as CCP) we investigated the pulsational behavior of cluster variables by assuming two different stellar masses for properly covering the range of pulsator masses expected in clusters with "solar scaled" metallicities $Z \simeq 0.0001$ ($M = 0.75 M_\odot$) and $Z \simeq 0.001$ ($M = 0.65 M_\odot$). In order to account for the suggested overabundance of alpha-elements, which would produce an increase in the "effective metallicity" (Straniero & Chieffi 1991) and a further decrease in the pulsator masses, a set of models characterized by a smaller mass value, $M = 0.58 M_\odot$, was also computed. According to CCP, this pulsator mass corresponds to a cluster effective metallicity $Z \simeq 0.006$. This value can be obtained by increasing, in a solar scaled metallicity of Z=0.001, the amount of alpha-elements of a factor roughly equal to 9 (see Salaris, Chieffi & Straniero 1991).

In the following section we provide an atlas of selected theoretical light curves arranged according to the effective temperature and the luminosity of each model. In §3 nonlinear periods are compared to the periods obtained by using the van Albada and Baker (1971) relations. In order to accomplish a comparison with observational data from RR Lyrae stars belonging to GGCs, the results of pulsation theory will be connected with the canonical HB evolutionary scenario.

Section 4 deals with the pulsational amplitude as a function of effective temperature, luminosity and mass. These results are collected in §5 for providing a theoretical Bailey diagram which is examined in terms of evolutionary expectations. In §6 this theoretical scenario is compared with observational evidences from RR Lyrae pulsators in selected GGCs. A final discussion about some residual but relevant mismatches between theory and observations closes the paper.

## 2. Light Curves

For each given value of the stellar mass the modal stability of the envelope models has been explored at selected luminosity levels ($log L/L_\odot = 1.51, 1.61, 1.72, 1.81, 1.91$) and by assuming a step of 100 K in the effective temperature.

Figure 1 shows the topology of the instability strip for $M=0.58\,M_\odot$, as compared with similar boundaries but for the mass values $M = 0.65$ or $0.75 M_\odot$ already reported in Bono, Caputo & Marconi (1995). Table 1 gives luminosities and effective temperatures along the newly computed boundaries. One finds that the topology of the instability strip for the mass value $M = 0.58 M_\odot$ discloses the same overall features of the instability strip of RR Lyrae models characterized by larger mass values. From the same figure it can be seen quite clearly that the width of the region where only the FO presents a stable limit cycle increases as the stellar mass increases.

Figure 1 also shows the location of some FO models for $M = 0.58 M_\odot$ which present a stable limit cycle at effective temperatures lower than the FO red edge. These models define a FO *stability isle* in a region of the instability strip where only F pulsators should be characterized by a stable limit cycle.

A similar but specular finding was obtained by BS. Indeed, they found some F models showing a stable limit cycle in the region of the instability strip where only FO pulsators might be present. The approach to limit cycle stability of these models due to the peculiarities of the dynamical and convective structure will be discussed in a forthcoming paper (Bono, Marconi & Stellingwerf 1996).

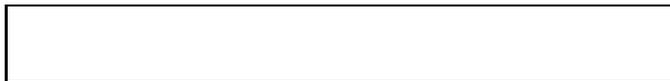

**Fig. 1.** The location into the HR diagram of the instability strip for $M = 0.58 M_\odot$ is compared with previous results for larger values of stellar mass and fixed chemical composition. At the lower luminosities, and for each assumed mass value, moving from higher to lower effective temperatures the different curves show: the FO blue edge (FO-BE), the F blue edge (F-BE), the FO red edge (FO-RE) and the F red edge (F-RE). Asterisks mark the location of some models for $M = 0.58 M_\odot$ which are characterized by a stable first overtone limit cycle in the region of the instability strip where only fundamental pulsators should be present. See text for further details.

**Table 1.** Fundamental and first overtone instability boundaries for $M = 0.58 M_\odot$.

| $\log L/L_\odot$ | FO-BE (°K) | F-BE (°K) | FO-RE (°K) | F-RE (°K) |
|---|---|---|---|---|
| 1.86 | 6850 |      | 6150 |      |
| 1.81 | 6950 | 7050 | 6750 | 5750 |
| 1.77 | 7050 | 7050 | 6750 |      |
| 1.72 | 7150 |      | 6650 |      |
| 1.61 | 7250 | 7150 | 6750 | 5950 |

In order to achieve a good accuracy throughout all phases of the pulsation cycle, a variable time step has been adopted: the number of time steps necessary for covering a period ranges roughly from 400 to 600. Before being able to provide any sound conclusion concerning the modal stability and the dependence of the light and velocity curve morphologies on physical parameters and chemical composition, the pulsation characteristics have to approach a limiting amplitude. To accomplish this goal the local conservation equations and the convective transport equation for each case were integrated in time until the initial velocity perturbation settles down and the pulsation behavior (amplitudes, periods) approaches a limit cycle stability.

The time interval necessary for the radial motions to approach their asymptotic amplitudes ranges from one thousand to ten thousand pulsational cycles. As a general rule, we assumed the direct time integration to be roughly equivalent to the inverse of the linear growth rate. Therefore, the nonlinear unstable models of the present survey are all characterized, over two consecutive periods, by a periodic similarity of the order of or lower than $10^{-(4 \div 5)}$. The range of luminosities covered by the different series of models was chosen so as to largely cover the theoretical prescriptions from HB evolutionary models. For variable stars these models foresee a minimum Zero Age Horizontal Branch (ZAHB) luminosity $\log L_{ZAHB} \simeq 1.70$ for $Z=0.0001$ and $\log L_{ZAHB} \simeq 1.65$ for $Z=0.001$ respectively. At the same time, the canonical HB evolutionary scenario predicts the occurrence of variable stars which are evolving within the instability strip at moderately larger luminosities. Figure 2 presents an atlas of theoretical light curves of F pulsators for two consecutive periods. These models were computed by assuming a mass $M=0.65\,M_\odot$. The light curves of FO pulsators referred to the same sequences of models are reported in Figure 3. Figures 4 and 5 show the same quantities of Figure 2 and 3 but are referred to the sequences in which a mass value $M=0.75\,M_\odot$ was adopted. Both light and velocity curves for the models discussed in this paper are available upon request to the authors.

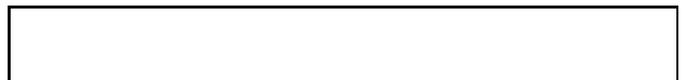

**Fig. 2.** Theoretical light curves of the four sequences of fundamental pulsators for the stellar mass $M=0.65\,M_\odot$. Each plot shows the bolometric amplitude for two consecutive periods and the effective temperature of the model. The luminosity level of the four sequences are also reported.

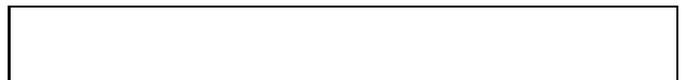

**Fig. 3.** Same as Fig. 2, but the light curves are referred to the four sequences of first overtone pulsators.

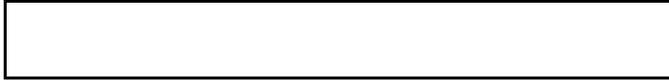

**Fig. 4.** Theoretical light curves of the four sequences of fundamental pulsators for the stellar mass M=0.75 M$_\odot$. Each plot shows the bolometric amplitude for two consecutive periods and the effective temperature of the model. The luminosity level of the four sequences are also reported.

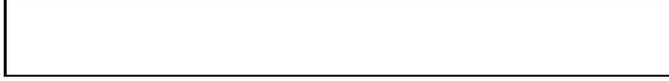

**Fig. 5.** Same as Fig. 4, but the light curves are referred to the four sequences of first overtone pulsators.

From the calculated light curves we find that the overall morphology, as well as the occurrence of secondary features like "Bumps" or "Dips" over the pulsation cycle closely follows the theoretical scenario suggested by BS which is based on models computed by adopting a different helium abundance (Y=0.30). The overall agreement concerning the occurrence of secondary features both in theoretical and observational light curves has been already discussed by Walker (1994) by presenting the photometry of RR Lyrae stars in the galactic globular M68.

## 3. Nonlinear periods

Figures 6 and 7 show the distribution of periods for all computed models with mass M=0.65 and 0.75 $M_\odot$ which reach a stable limit cycle either in the F or in the FO mode. In the same figures we report the linear predictions given by van Albada & Baker (1971) according to the relations:

$$\log P_F = 11.497 + 0.84 \log L - 0.68 \log M - 3.48 \log T_e$$
$$\log \left(\frac{P_F}{P_{FO}}\right) = 0.43 + 0.14 \log L - 0.032 \log M - 0.009 \log T_e$$

where both mass and luminosity are in solar units, whereas the periods are in days.

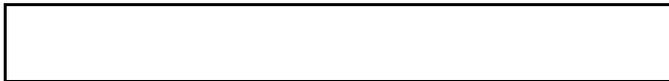

**Fig. 6.** Fundamental periods versus effective temperatures for two different stellar masses $M/M_\odot$=0.65 (open circles) and $M/M_\odot$=0.65 (solid circles). Solid lines refer to the nonlinear results of the present computations, whereas dotted lines show the linear periods evaluated by using the relations provided by van Albada & Baker (1971). The luminosity level of each sequence is also reported at the right of the curves.

As expected, nonlinear periods are generally smaller than linear periods, since convective motions smooth the density profile in coincidence of the hydrogen ionization region where the density inversion takes place, and in the mean time produce a non negligible increase of the density in the region where the adiabatic exponent -$\Gamma_1$- reaches its minimum value. The aftermaths of this effect together with the nonlinear effects on $\Gamma_1$ cause the decrease of the nonlinear periods when compared to linear ones. The previous leading-term considerations also provide an explanation of the physical reason why this discrepancy tends to increase moving from the blue to the red side of the instability strip (see Figure 11 and 12 in BS).

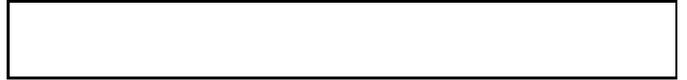

**Fig. 7.** Same as Fig. 6, but for first overtone pulsators.

The new models show that the differences are larger for FO than for F pulsators, whereas BS found that the discrepancy between linear and nonlinear periods was larger for F pulsators. Since the present nonlinear models differ from the models adopted in BS only in the helium abundance, this finding strongly suggests that the FO periods present a non negligible dependence on this parameter. Linear interpolation among the data plotted in both figures gives the new relations:

$$\log P_F = 11.627 + 0.823 \log L - 0.582 \log M - 3.506 \log T_e$$
$$\log P_{FO} = 10.789 + 0.800 \log L - 0.594 \log M - 3.309 \log T_e$$

with a correlation coefficient r=1.00. The differences between the previous relations and the van Albada & Baker relations -though not negligible- are not expected to produce a substantial change in the available pulsational scenario, supporting once more the pioneering work by van Albada & Baker. However, previous results give at least a firm warning against the long-standing attempt to derive the mass of double-mode pulsators by using the ratio between F and FO periods ($P_F/P_{FO}$). Indeed, this quantity critically depends on both stellar parameters and on even minor details of the physical and numerical ingredients adopted in the pulsational codes. The occurrence and the properties of RR Lyrae models which show stable double-mode limit cycles will be discussed in a forthcoming paper.

The previous relations, which provide the nonlinear periods for the first two pulsating modes, can be combined with the instability strip topologies given by BCM to derive the behavior of the minimum period allowed in a cluster, i. e. the minimum (fundamentalized) period of the FO pulsator at the blue edge of its instability region:

$$\log P_{FO-BE} = -2.424 - 0.714 \log M + 1.046 \log L$$

which appears in good agreement with the preliminary evaluation obtained by Bono et al. (1995). As discussed in the quoted paper the minimum periods provided by the previous relation reproduce quite satisfactorily the minimum periods of variable stars belonging to GGCs, thus supporting once more the canonical evolutionary scenario against the suggested occurrence of an anomalous period shift.

As a further step in our investigation, let us take now into account the canonical evolutionary constraints given by CCP. Figure 8 shows the expected location of HB evolutionary tracks for selected values of the cluster metallicity. This parameter was chosen such that it should become representative of well known clusters which populate the galactic halo and present a large number of RR Lyrae variables, namely Z=0.0001 (M15), Z=0.0004 (M3) and Z=0.001 (M5).

In the same figure we plot the instability strip for the two labeled assumptions about the mass of the star, giving, for each assumed metallicity the maximum mass value ($M_{15}^{max}$) allowed by evolutionary models for populating the HB of a cluster characterized by an age of 15 Gyr and the expected minimum period of $ab$ type pulsators (see below). It appears that the instability strip in M15 should be populated by stars with mass values included between 0.75 and 0.80 $M_\odot$. More massive stars would be forbidden for the adopted cluster age, thus giving a straightforward explanation of the blue HB shown by this cluster.

On the contrary, the instability regions of M3 and M5 should be populated by stars with mass values which range from 0.65 to 0.70 $M_\odot$ respectively. These values imply now an amount of mass loss of the order of 0.10 $M_\odot$ or 0.18 $M_\odot$ for M3 and M5 respectively.

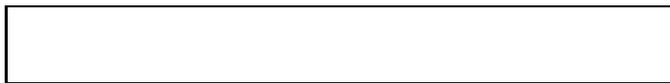

**Fig. 8.** The ZAHB location into the HR diagram for three different values of the cluster metallicity. In each plot the solid lines show the evolutionary paths of selected helium burning models for the labeled values of the masses. Dashed lines and dotted lines report the instability strip for two different mass values M=0.75 $M_\odot$ and M=0.65 $M_\odot$ respectively. For each given metallicity we also report the name of the typical cluster, the observed minimum period of fundamental pulsators in that cluster ($logP_{ab}^{min}$), and the maximum mass allowed for populating the HB ($M_{15}^{max}$). See text for further explanations.

As a most relevant point, we find that this evolutionary picture can be easily connected with the observed minimum period of $RR_{ab}$ pulsators, i.e. with the observational evidence at the basis of the Oosterhoff dichotomy. The first suggestion about the occurrence of the Oosterhoff dichotomy was produced by the observational evidence that the mean periods of F pulsators in GGCs, $<P_{ab}>$, were grouped around two distinct values, namely 0.65 d (OoII) and 0.55 d (OoI), with a lack of clusters with $<P_{ab}>$=0.6 d. As it was early recognized, such an occurrence is due to the fact that when compared to OoII clusters, OoI clusters allow the occurrence of F pulsators characterized by shorter periods. Indeed, the minimum F period changes from $\log P_{ab}^{min} \simeq$ -0.25 in OoII clusters to $\log P_{ab}^{min} \simeq$ -0.35 in OoI clusters.

This interesting observational behavior has stimulated an animated debate during the past few years. According to Sandage (1958, 1981, 1982, 1990) the difference in periods should mainly be ascribed to a difference in luminosity: OoII pulsators are more luminous than OoI ones.

In order to explain the F mean period distribution, van Albada & Baker (1973) suggested an alternative hypothesis: a difference in the maximum temperature of F pulsators produced by a hysteresis mechanism. In this context a variable star located in the OR region will pulsate in the F mode if it is evolving from lower to higher effective temperatures, whereas it will pulsate in the FO if the star is evolving in the opposite direction. As a consequence of this mechanism, we expect that the transition between $RR_{ab}$ and $RR_c$ occurs either at the F blue edge or at the FO red edge.

In the already quoted Figure 8, the point where the transition is expected to take place according to the hysteresis hypothesis is marked with an open circle. Accordingly one expects that the OR region is populated by FO pulsators in OoII clusters, and by F pulsators in the OoI case. According to such an hypothesis, the data listed in Table 2 disclose that the theoretical results concerning $P_{ab}^{min}$ overlap surprisingly well with the observational data when both evolutionary prescriptions and pulsational results are taken into account.

If one adds the evidence given in BCM for a HR distribution in agreement with the hysteresis prescriptions, we can conclude that the canonical evolutionary scenario appears able to account for several relevant features of RR Lyrae pulsators, without the need of invoking "ad hoc" modifications of the current evolutionary knowledge.

**Table 2.** Theoretically predicted minima of fundamental periods as a function of cluster metallicity.

| Z | M | $\log L$ | $\log T_e$ | $\log P_{ab}^{min}$ | Transition |
|---|---|---|---|---|---|
| 0.0001 | 0.80 | 1.761 | 3.820 | -0.26 | FO-RE |
| 0.0004 | 0.70 | 1.710 | 3.844 | -0.35 | F-BE |
| 0.0010 | 0.65 | 1.659 | 3.843 | -0.37 | F-BE |

As a final point, let us note that the data in Table 2 give a spontaneous indication for the shift in periods *within* the OoI type clusters (Castellani & Quarta 1987), with F pulsators in M5 reaching shorter periods with respect to the same type of pulsators in M3.

## 4. Pulsational amplitudes

A first attempt to connect pulsational amplitudes with stellar parameters has been recently presented by Brocato, Castellani & Ripepi (1996) who discussed new data for RR Lyrae stars in the globular M5. However, their investigation was necessarily based on the available pulsational computations given by BS for a fixed but unrealistic amount of atmospheric helium (Y=0.30). Similar pioneering results are now being compared with amplitudes com-puted under the much more realistic assumption Y=0.24 (Pagel 1995, Peimbert 1995).

Figure 9 shows the bolometric amplitude as function of the effective temperature for a star of 0.65 $M_\odot$ at selected stellar luminosity values and for the two labeled assumptions of atmospheric helium Y=0.24 or Y=0.30. Note that the figure reports only the amplitudes of models which present a stable limit cycle in the F and/or in the FO. One finds that the relation between period and temperature obtained by assuming a helium content $Y = 0.24$ does not differ considerably from that obtained by assuming a larger helium abundance ($Y = 0.30$). The small differences in the range of temperatures taken into account appear as the direct consequence of the different location in the HR diagram of the corresponding instability regions.

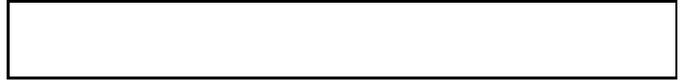

**Fig. 10.** Bolometric amplitude versus effective temperature for both fundamental and first overtone variables. The amplitudes plotted in this figure are referred to models computed by assuming a fixed helium content (Y=0.24) and two different stellar masses M=0.65 $M_\odot$ (solid lines), and M=0.75 $M_\odot$ (dashed lines). The sequences of models characterized by different luminosity levels are marked with different symbols.

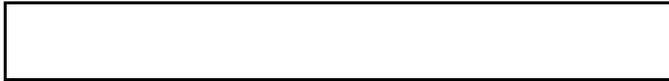

**Fig. 9.** Bolometric amplitude versus effective temperature for both fundamental and first overtone variables. The amplitudes plotted in this figure are referred to models computed by assuming a fixed stellar mass M=0.65 $M_\odot$ and two different helium abundances Y=0.24 (solid lines) and Y=0.30 (dashed lines). The sequences of models characterized by different luminosity levels are marked with different symbols.

A relevant feature for Y=0.24 is that the correlation between the amplitudes of F pulsators and the effective temperature becomes more and more independent of stellar luminosity in comparison with the models which present a higher helium abundance (Y=0.30). This is of course a crucial point, since it leads to a direct correlation between F-amplitudes and effective temperatures, as earlier suggested by Sandage (1981). Moreover, one finds that the distributions of FO amplitudes versus effective temperature keeps showing the characteristic "bell" shape, with the maximum amplitude sensitively dependent on the luminosity level of the pulsators. These amplitudes moving from Y=0.30 to 0.24 do not show systematic differences, since at higher luminosities the two curves are almost identical. At lower luminosities the models with Y=0.30 present pulsational amplitudes which are initially larger and then smaller in comparison with the models characterized by Y=0.24.

Before attempting to match pulsational and evolutionary prescriptions, it is necessary to evaluate the effects of varying the stellar mass in the above quoted scenario. This is shown in Figure 10 in which we report pulsational amplitudes against effective temperatures but for fixed helium content (Y=0.24) and for two different assumptions on stellar mass M=0.65 and 0.75 $M_\odot$.

Figure 10 shows that a mass increase has little effect on the amplitudes of F-pulsators, whose independence of the luminosity level appears even strengthened for higher masses. On the contrary, at lower luminosities we find that an increase of the stellar mass implies an increase in FO amplitudes. As a point that will be relevant further on,

Figure 10 supports the suggestion of BS that the amplitude of FO pulsators along the decreasing branch leading to the red limit is largely independent of the stellar luminosity. We further find that, for a given temperature, this amplitude appears moderately dependent on the pulsator mass, increasing when the mass increases. We can find a reason for such a behavior by recognizing that the decreasing branch appears populated by FO pulsators in the OR-region, and by assuming that the pulsational amplitude only depends on the distance in temperature from the red edge. According to such a picture, FO amplitudes are expected to be largely independent of the luminosity for a given mass, while increasing the mass at any given temperature increases the amplitude because the FO boundary shifts toward cooler temperatures, thus increasing the distance of the given temperature from the FO-RE.

Concerning the predicted behavior of stars in GGCs, one may notice that the amplitude distributions shown in Figure 10 give two different periods in the OR-region, one along the F and the other along the FO sequence. If the suggestion advanced in the previous section about the efficiency of the hysteresis mechanism is correct, one expects OR-regions populated according to the Oosterhoff type, either by FO (OoII) or by F pulsators (OoI). In terms of the data shown in Figure 10 one expects in OoII type clusters the occurrence of a well developed decreasing branch of FO pulsators together with a sequence of F pulsators depopulated in its upper portion. On the contrary, in OoI type clusters one expects a sequence of F pulsators populated up to the largest amplitude, with negligible evidence for the decreasing branch of first overtones.

The discussed scenario can be safely extended to the case for M=0.58 $M_\odot$, as illustrated by data in Figure 11. Indeed, both the FO amplitudes and the width in temperature of the FO region regularly decrease when passing from M=0.75 $M_\odot$ to 0.65 and 0.58 $M_\odot$. Moreover, one finds that F pulsators arrange even better along the previous amplitude-temperature relation. As a result, if pulsational prescriptions given through figures 10 and 11 are taken at their face values, the observed amplitudes of F pulsators should allow the determination of stellar effective temperatures with an error of about ±100 K, if not better, independently of the stellar mass and luminosity.

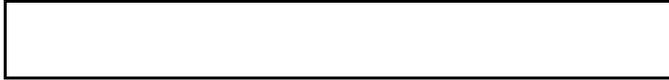

**Fig. 11.** Pulsational amplitudes versus effective temperature for three different assumptions on stellar mass. First overtone amplitudes refer to the common value of luminosity $\log L/L_\odot = 1.61$. Fundamental amplitudes are given for $\log L/L_\odot = 1.61$, 1.81 (M=0.58 $M_\odot$) and $\log L/L_\odot = 1.72$, 1.81. 1.91 (M=0.65 $M_\odot$).

## 5. The theoretical Bailey period-amplitude diagram

Theoretical periods and theoretical amplitudes, as obtained for each given assumption about stellar mass, can be finally collected to produce a theoretical period-amplitude diagram, thus allowing to investigate the connection of these two pulsational parameters with the evolutionary parameters. Figure 12 shows the distributions into this theoretical Bailey diagram when the stellar temperature is varied across the instability region, for selected values of the stellar luminosity and for the two choices M=0.65 $M_\odot$ (solid lines) and M=0.75 $M_\odot$ (dashed lines), keeping everywhere $Y = 0.24$. The pulsational amplitudes plotted in this figure are referred to four different luminosity levels and cover the overall instability region.

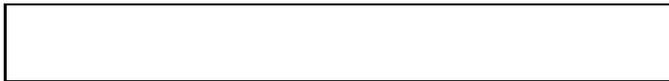

**Fig. 12.** Theoretical Bailey diagram, bolometric amplitude versus the logarithm of the period for a fixed helium content (Y=0.24) and two different mass values M=0.65 $M_\odot$ (solid lines), M=0.75 $M_\odot$ (dashed lines). Symbols concerning the luminosity levels are the same as in Fig. 10.

Bearing in mind the distribution reported in Figures 10 and 11, the topology of data plotted in Figure 12 can be easily understood in terms of the relation which connects the periods with effective temperatures, luminosities and masses. By using these relations, for a given value of the mass and for a fixed luminosity level, the period scales according to $\log P \propto 3.5$ (or 3.3) $\log Te$ and the distributions shown in Figure 10 can be therefore translated into the distributions shown in Figure 12. Moreover, by increasing the luminosity level the periods increase by about $\delta \log P \simeq 0.8 \log L$. Such an occurrence removes the "degeneracy" of F amplitudes with stellar luminosity and produces the period shift observed in Figure 12 for a fixed value of the mass. According to these simple arguments, we finally find that the amplitude distribution, by increasing the mass, moves toward shorter periods, thus giving full account of the Bailey diagram morphology disclosed in Figure 12.
Adopting such a pulsational scenario, it is obviously interesting to match pulsational constraints with the predictions of the evolutionary theory. For a quick look on this new scenario let us first discuss the distribution of HB stars in the Bailey diagram *by assuming that all stars are just in their ZAHB location.* Under this assumption, for each given value of the metallicity we derived the set of stellar masses populating the instability strip by interpolating the evolutionary tracks given by CCP, each mass being characterized by proper values of both effective temperature and luminosity. The regular behavior of F pulsators allows a straightforward linear interpolation of the dependence of the amplitude on stellar mass, temperature and luminosity. On this basis evolutionary data can be easily transformed into the Bailey diagram for F pulsators as given in Figure 13 for the three different choices of metal abundance Z=0.0001, Z=0.0004 and Z=0.001.
The procedure previously outlined supplies a relevant result that can be summarized as follow:
*the difference of the evolutionary structure of the pulsators predicted on the basis of different assumptions about the stellar metallicity plays a minor role in the distribution of the pulsators in the Bailey diagram.*
As suggested by Brocato, Castellani & Ripepi (1996) on a pure observational basis, this is due to the fact that moving from metal poor OoII to metal rich OoI pulsators the decrease in mass is largely balanced by the decrease in luminosity as can be seen in Figure 8.
Unfortunately, the distribution of FO pulsators appears much less predictable. Due to the rather intricate amplitude behaviors disclosed in Figures 10 and 12, interpolation over the whole range of theoretical data appears not every useful, until a much finer set of nonlinear luminosity amplitudes becomes available. However, the regular behavior of the decreasing branch allows again a linear interpolation of the dependence of such a feature on stellar parameters. On this basis the distribution of FO pulsators could be predicted.
We eventually underline that the distribution of F pulsators in the Bailey diagram, even when taking into account the off ZAHB evolution, would be affected only marginally, since these variables are characterized by a well defined amplitude-temperature relation which presents a negligible dependence on both stellar mass and luminosity level (see Figure 9, 10 and 11). The conclusions previously drawn for F pulsators cannot be extended to FO pulsators due to the strong dependency of their pulsational amplitudes on luminosity level. As a consequence the distribution of these variables in the Bailey diagram, with the exception of pulsators located along the decreasing branch, might be partially changed by the off ZAHB evolution.

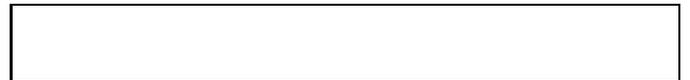

**Fig. 13.** The distribution in the theoretical Bailey diagram of ZAHB fundamental and first overtone pulsators as expected according to evolutionary prescriptions for the labeled assumptions about the stellar metallicity. The distribution of the decreasing branch of first overtone pulsators is also sketched. See text for further explanations.

## 6. Comparison with observations

To approach the problem of comparing the present theoretical scenario with observational data, in Figure 14 we show the observational Bailey diagram given by Sandage (1990, and references therein) for pulsators belonging to the two well studied clusters M15 (OoII) and M3 (OoI). A quick look to the distributions plotted in this Figure reveals a series of relevant similarities with theoretical results that are worth listing:

i) F pulsators in both OoI and OoII clusters arrange along a rather continuous sequence.

ii) The OoII prototype M15 lacks F pulsators in the upper portion of the amplitude distribution, whereas a large amount of M3 (OoI prototype) $RR_{ab}$ variables populate this region.

iii) The lack of large amplitude F pulsators in M15 is accompanied by the clear evidence for a decreasing branch of FO pulsators. On the contrary, M3 shows a much less clear evidence of such a decreasing branch, if any.

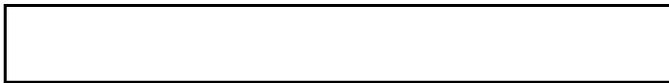

**Fig. 14.** The observational Bailey diagram, blue amplitude versus the logarithm of the period, for RR Lyrae variables belonging to galactic globulars M3 (solid circles) and M15 (open circles) as given by Sandage (1990).

According to previous findings, we can conclude that the proposed theoretical scenario offers an interesting clue to an evolutionary interpretation of the observed features shown by the Bailey diagram supporting the occurrence of F (or FO) pulsators in OoI (OoII) clusters. However, a more detailed comparison reveals several quantitative mismatches which will be discussed in detail.

To allow such a quantitative comparison with actual cluster pulsators, theoretical light curves have been transformed through Kurucz's (1992) evaluation of bolometric corrections and color-temperature relations to derive amplitudes in the B photometric band, which is the most adopted observational parameter for RR Lyrae amplitudes. As it is well known, when the luminosity of the pulsator decreases the star becomes cooler and the bolometric correction increases. As a consequence, B amplitudes for RR Lyrae pulsators are expected to be larger than the corresponding bolometric amplitudes, an occurrence which of course justifies the widespread use of such an observational parameter. Panel a and b of Figure 15 show the predicted amplitude-temperature and amplitude-period relations transformed in terms of B magnitudes.

On the basis of this new pulsational topology the analysis given in Figure 13 can now be repeated for providing theoretical predictions to be compared with observations. We report in Table 3 for the metal content Z=0.0001 the expected pulsational properties of F and FO pulsators along the ZAHB.

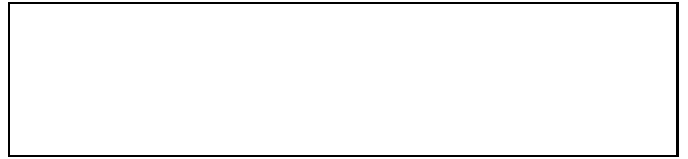

**Fig. 15.** The theoretical (a) amplitude-temperature and (b) amplitude- period relations for RR Lyrae variables as given in terms of blue amplitudes A(B).

**Table 3.** Predicted ZAHB fundamental and first overtone pulsators for Z=0.0001 (mass and luminosity are in solar units).

| $M$ | $\log L$ | $\log T_e$ | $\log P_F$ | $A_F(B)$ mag. | $\log P_{FO}$ | $A_{FO}(B)$ mag. |
|---|---|---|---|---|---|---|
| 0.795 | 1.715 | 3.835 | -0.350 | 1.982 | -0.470 | 1.212 |
| 0.800 | 1.718 | 3.830 | -0.331 | 1.835 | -0.452 | 1.029 |
| 0.810 | 1.724 | 3.821 | -0.296 | 1.568 | -0.421 | 0.700 |
| 0.820 | 1.729 | 3.812 | -0.266 | 1.339 | -0.390 | 0.370 |
| 0.830 | 1.734 | 3.806 | -0.242 | 1.146 | -0.370 | 0.154 |
| 0.840 | 1.737 | 3.800 | -0.223 | 0.991 | | |
| 0.850 | 1.740 | 3.796 | -0.209 | 0.872 | | |

If we take into account only the decreasing branch of the FO amplitudes displayed in the left panel of Figure 15, for not too large luminosities, we derive the following relation:

$$A(B) = 37.73 \log T_e + 1.93 \log M - 143.29$$

On the basis of such a relation we can attempt a prediction for variable stars located in the OR region, i.e. for stars already evaluated as F pulsators at the top of the F-branch. Data for these FO pulsators are also reported in Table 3.

The comparison with observational data for pulsators belonging to M15 is shown in Figure 16. In the same figure are also plotted the evolutionary paths of F pulsators evolving off their ZAHB loci. Comparison with data for M15 discloses a surprising agreement. However, the same figure shows that theory foresees the occurrence of large amplitudes (A(B)$\simeq$1.1 mag) for FO pulsators which are not present in the cluster.

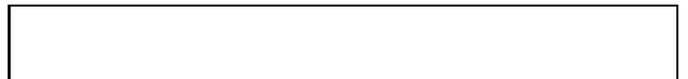

**Fig. 16.** Observational data for M15 pulsators are compared with the theoretical distribution for Z=0.0001 ZAHB pulsators. In the same figure are also shown the evolutionary paths for the labeled values of stellar masses.

The procedure has been repeated adopting Z=0.0004, assumed as a suitable choice for OoI clusters like M3. Data for both F and FO pulsators are reported in Table 4, whereas in Figure 17 theoretical predictions are compared with observational data for the quoted clusters. The agreement appears now somehow less satisfactory, since some F pulsators tend to have lower periods than predicted by the theory. If this occurrence is not an artifact of forcing the results for the lower luminosity levels into a common linear interpolation, we find no obvious origin for such a behavior.

**Table 4.** Predicted ZAHB fundamental and first overtone pulsators for Z=0.0004.

| $M$ | $\log L$ | $\log T_e$ | $\log P_F$ | $A_F(B)$ mag. | $\log P_{FO}$ | $A_{FO}(B)$ mag. |
|---|---|---|---|---|---|---|
| 0.690 | 1.671 | 3.833 | -0.342 | 1.820 | -0.462 | 1.018 |
| 0.700 | 1.682 | 3.810 | -0.257 | 1.230 | -0.381 | 0.162 |
| 0.705 | 1.687 | 3.800 | -0.218 | 0.973 | | |
| 0.710 | 1.693 | 3.791 | -0.183 | 0.711 | | |
| 0.715 | 1.699 | 3.782 | -0.150 | 0.472 | | |
| 0.720 | 1.704 | 3.774 | -0.121 | 0.257 | | |
| 0.725 | 1.708 | 3.768 | -0.095 | 0.065 | | |

As for F pulsators once again we find that theoretical amplitudes for FO pulsators are larger than observed values. However, we look at the comparison in Figure 17 as an evidence of the lack of the decreasing branch, disregarding the small amplitude variations at the larger periods as an evidence for the present location of the decreasing branch phase.

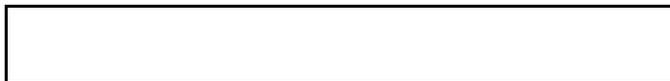

**Fig. 17.** Observational data for M3 pulsators are compared with the theoretical distribution for Z=0.0004 ZAHB pulsators. In the same figure are also shown the evolutionary paths for the labeled values of stellar masses.

As a final point, in Figure 18 are compared theoretical and observational data concerning the dependence of blue amplitudes on stellar effective temperatures. According to Brocato, Castellani & Ripepi (1996) the pulsator mean colors have been first corrected for reddening and then transformed into the true colors of the static models, using the procedure given by Bono, Caputo & Stellingwerf (1995). Temperatures have been finally derived adopting the color-temperature relation as given by Kurucz(1992). We find a further disagreement between observations and theory. Indeed observational amplitudes appear smaller for each given value of the effective temperature or, conversely, for each given amplitude the temperature appears about 300 K larger than predicted by the theory. Note that data in Figure 18 assume a reddening value E(B-V)=0.07 for M15 (Zinn 1985). By adopting the reddening value suggested by Sandage (1990), i. e. E(B-V)=0.11, the discrepancy for this cluster further increases with 200 K. The reader is referred to Brocato, Castellani & Ripepi (1996) and to Silbermann & Smith (1995) for a detailed discussion concerning the M15 reddening evaluations.

As a conclusion, the theoretical pulsational scenario we are dealing with, although it presents relevant similarities with the actual behavior of cluster pulsators, is far from reaching a tight correspondence with the available observational data.

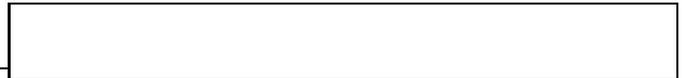

**Fig. 18.** Theoretical expectations concerning the B amplitude-temperature relations for Z=0.0001 or Z=0.0004 are compared with observational data for the cluster M15 and M3, respectively. Equilibrium temperatures are from Bono, Caputo & Stellingwerf (1995).

One could take the comparison in Figure 18 as an evidence that our theoretical approach could be affected by a systematic overestimate of amplitudes. As a matter of fact, the difference in temperature of about 300 K for a given amplitude appears much larger than the usual estimates of errors in the color-temperature relation of pulsating stars. If this is the case, the surprising agreement we found for the Bailey diagram of metal poor clusters should be an artifact of balancing errors.

However, in order to properly set the present observational scenario concerning the systematic errors in the RR Lyrae effective temperature scale, it is worth mentioning that, by adopting the V-R color-temperature relation provided by Vanderberg & Bell (1985) instead of Kurucz's relation, Silbermann & Smith (1995) obtained a systematic shift toward cooler temperatures. Most interesting, they also found that by adopting the temperature calibration derived by Longmore et al. (1990) on the basis of V-K colors, the RR Lyrae temperatures are in average 300 K cooler than the temperature obtained by using the V-R colors. Here we can only present this contradictory scenario as something to be explored before reaching the final goal of reading the location of RR Lyrae pulsators in the Bailey diagram in terms of the evolutionary status of cluster stars.

## 7. Discussion and conclusions

Before closing the argument, let us discuss if and how the theoretical scenario could be varied in the attempt to match the theoretical results with observations. The theoretical framework outlined in the present paper is based on numerical experiments supplied by three different fields: 1) stellar evolution, 2) stellar pulsation, 3) stellar atmospheres. Based on proper physical assumptions, all of them simulate the physical processes which take place in different regions of the stellar structure. Unfortunately, any observational test involves in the same context the results of the quoted scenario so that it appears difficult to point out where and why the theoretical outcomes could be affected by systematic errors. In order to give some more light on this scenario, let us briefly discuss in the following the three above quoted theoretical ingredients.

1) Several and thorough papers have been recently devoted to the evolutionary properties of HB stars and to their dependence on astrophysical parameters and physical ingredients (CCP; Dorman, Rood & O'Connell 1994 and references therein) and therefore they will not be discussed here.

2) The theoretical treatment of radial stellar pulsations is based on two main ingredients: the physical inputs necessary to produce an envelope model and the set of convective hydrodynamic equations adopted for following the nonlinear coupling between dynamical and convective motions.

Among the physical inputs, both radiative opacities and equations of state are worth being checked, since the radial pulsational instability is driven by the destabilizing effects operated by *kappa* and *gamma* mechanisms. Numerical tests performed to verify the dependence of the nonlinear pulsation characteristics on the new OP (Seaton et al. 1994) and OPAL (Roger & Iglesias 1992) opacities show almost no difference in comparison with models based on the old Los Alamos opacities. A similar result was obtained by Buchler & Buchler (1994) in their investigation on BL Herculis variables. Therefore this ingredient, at least for low metallicity stars, has no effect on the amplitude problem.

Concerning the equation of state we did not perform any specific calculation since recent nonlinear results on radiative models of bump Cepheids provided by Kanbur (1992) show only negligible physical differences on the adopted equations of state. However, we plan to test the new Livermore equation of state as soon as it becomes available for metal contents suitable to metal poor stars.

Although the assumptions we adopted to describe the turbulent field in variable stars are physically plausible, the convective transport equation has been derived by using three different free parameters (for a detailed discussion see Stellingwerf 1982 and BS). A more formal and rigorous approach for deriving the set of radial pulsation equations was adopted by Kuhfuß (1986) and by Gehmeyr (1992 and references therein). However, only few full amplitude nonlinear models have been computed by adopting the latter formalism and therefore a trustworthy comparison between hydrocodes which involve different physical and numerical approximations cannot be provided.

The self-consistent inclusion of nonlocal, nonlinear and time-dependent effects may help to improve long-standing pulsation questions (BS; BCM; Bono, Caputo & Stellingwerf 1995). In particular, the nonlinear convective models are characterized by pulsation amplitudes that, although still too large, are roughly half magnitude smaller than the amplitudes of radiative models.

3) The transformation of the light curves into the observational plane is one of the most thorny problems of radial stellar pulsations. At present two different routes can be followed to accomplish this transformation. The former involves the calculation of hydrodynamic pulsating atmosphere models which take into account a multifrequency radiation transport in the outermost optically thin layers. This approach not only provides useful insights on the dynamical properties of the surface regions but can also predict broad-band colors. Until now only few investigations have been devoted to soundly solve this problem for RR Lyrae variables (Keller & Mutschlecner 1970; Bendt & Davis 1971).

Davis & Cox (1980) following this route computed few models for evaluating the RR Lyrae mean colors but no theoretical study has been undertaken so far for estimating the RR Lyrae light curves in specific photometric bands.

The latter route relies on the use of bolometric corrections and color-temperature relations provided by static atmosphere models. This approach is based on two weak physical assumptions: the stellar effective temperature over the full cycle is derived by assuming that the surface zone is always in radiative equilibrium, whereas the surface gravity is evaluated by assuming both radiative and hydrostatic equilibria (Bono, Caputo & Stellingwerf 1994). Although the quoted approximations were proved to be suitable for evaluating the RR Lyrae mean colors (Bono, Caputo & Stellingwerf 1995) and for reproducing the ionization and excitation equilibrium conditions of metal lines close to the phase of minimum light (Clementini et al. 1995), the evaluation of pulsation amplitudes may require a more cautious analysis.

We conclude that the present theoretical scenario proves to be successful to shed light on many pulsation features of cluster variables and, in particular, on long-standing questions such as the transition between FO and F pulsators and the origin of the Oosterhoff dichotomy. However, predictions concerning the pulsation amplitudes are far from reaching the required compatibility with observational data. The possible origins of such a discrepancy were discussed, but firm conclusions have not been reached.

We can only draw attention on such a situation, waiting for further improvements either of the theory of radial pulsation or of the theory of stellar atmospheres and/or of stellar evolution which may succeed in reconciling theory with observation. This will reach the relevant goal of using the pulsational parameters of RR Lyrae variables to gain firm information on their evolutionary status and, in particular, on their masses, temperatures and luminosities. The use of these pulsators will allow us to solve many open questions concerning the evolution of our Galaxy and, in particular, they might become robust standard candles for mapping the distances inside the Galaxy and among the galaxies belonging to the Local Group.

*Acknowledgments.* It is a pleasure to thank P.A. Mazzali for useful conversations on stellar atmospheres. We wish also to acknowledge A. Balestra and C. Vuerli of the OAT Technologic Division for their warm support in computing facilities. We are also grateful to C. Aerts for his valuable comments as referee on an early draft of this paper which have improved its readability. This work was partially supported by MURST, CNR-GNA and ASI.

# References

Bailey, S. J. 1899, ApJ, 10, 255
Bailey, S. J. 1902, Ann. Harward Obs., 38, 132
Bendt, J. E. & Davis, C. G. 1971, ApJ, 169, 333
Bono, G., Caputo, F., Castellani, V. & Marconi, M. 1995, ApJ, 448, L115
Bono, G., Caputo, F. & Marconi, M. 1995, AJ, 110, 2365 (BCM)
Bono, G., Caputo, F. & Stellingwerf, R. F. 1994, ApJ, 432, L51
Bono, G., Caputo, F. & Stellingwerf, R. F. 1995, ApJS, 99, 263
Bono, G., Castellani, V. & Stellingwerf, R. F. 1995, ApJ, 445, L145
Bono, G., Marconi, M. & Stellingwerf, R. F. 1996, in preparation
Bono, G. & Stellingwerf, R. F. 1992, Mem. Soc. Astron. It., 63, 357
Bono, G. & Stellingwerf, R. F. 1994, ApJS, 93, 233 (BS)
Brocato, E., Castellani, V. & Ripepi, V. 1996, AJ, 111, 809
Buchler, J. R. & Buchler, N. E. G. 1994, A&A, 391, 736
Buzzoni, A., Fusi Pecci, F., Buonanno, R. & Corsi, G. E. 1983, A&A, 230, 315
Castellani, V., Chieffi, A. & Pulone, L. 1989, ApJ, 344, 239 (CCP)
Castellani, V. & Quarta, M. L. 1987, A&AS, 71, 1
Christy, R. F. 1966, ApJ, 144, 108
Clementini, G., Carretta, E., Gratton, R., Merighi, R., Mould, J. R. & McCarthy, J. K. 1995, AJ, 110, 2319
Davis, C. G. & Cox, A. N. 1980, in Current Problems in Stellar Pulsation Instabilities, eds. D. Fischel, J.R. Lesh, W.M. Sparks, (NASA TM 80625), 293
Dorman, B., Rood, R. T. & O'Connell, R. W. 1993, ApJ, 419, 596
Gehmeyr, M. 1992, ApJ, 399, 272
Kanbur, S. M. 1992, A&A, 259, 175
Keller, C. F. & Mutschlecner, J. P. 1970, ApJ, 161, 217
Kuhfuß, R. 1986, A&A, 160, 116
Kurucz, R. L. 1992, in IAU Symp. 149, The Stellar Populations of Galaxies, ed. B. Barbuy & A. Renzini, (Dordrecht:Kluwer), 225
Longmore, A. J., Dixon, R., Skillen, I., Jameson, R. F. & Fernley, J. A. 1990, MNRAS, 247, 684
Pagel, B. E. J. 1995, in The Light Elements Abundances, ed. P. Crane, (Berlin: Springer Verlag), 155
Peimbert, M. 1995, in The Light Elements Abundances, ed. P. Crane, (Berlin: Springer Verlag), 165
Peimbert, M. & Torres-Peimbert, S. 1976, ApJ, 203, 581
Rogers, F. J. & Iglesias, C. A. 1992, ApJS, 79, 507
Salaris, M., Chieffi, A. & Straniero, O. 1991, ApJ, 414, 580
Sandage, A. 1958, in Stellar Populations, ed. D.J.K. O'Connell, (Città del Vaticano: Vatican Observatory), 41
Sandage, A. 1981, ApJ, 248, 161
Sandage, A. 1982, ApJ, 252, 553
Sandage, A. 1990, ApJ, 350, 603
Seaton, M. J. 1994, MNRAS, 266, 805
Silbermann, N. A. & Smith, H. A. 1995, AJ, 110, 704
Stellingwerf, R. F. 1975, ApJ, 195, 441
Stellingwerf, R. F. 1982, ApJ, 262, 330
Straniero, O. & Chieffi, A. 1991, ApJS, 76, 525
van Albada, T. S. & Baker, N. 1971, ApJ, 169, 311
van Albada, T. S. & Baker, N. 1973, ApJ, 185, 477
Vanderberg, D. A. & Bell, R. A. 1985, ApJ, 185, 477
Walker, A. R. 1994, AJ, 108, 555
Zinn, R. 1985, ApJ, 293, 424

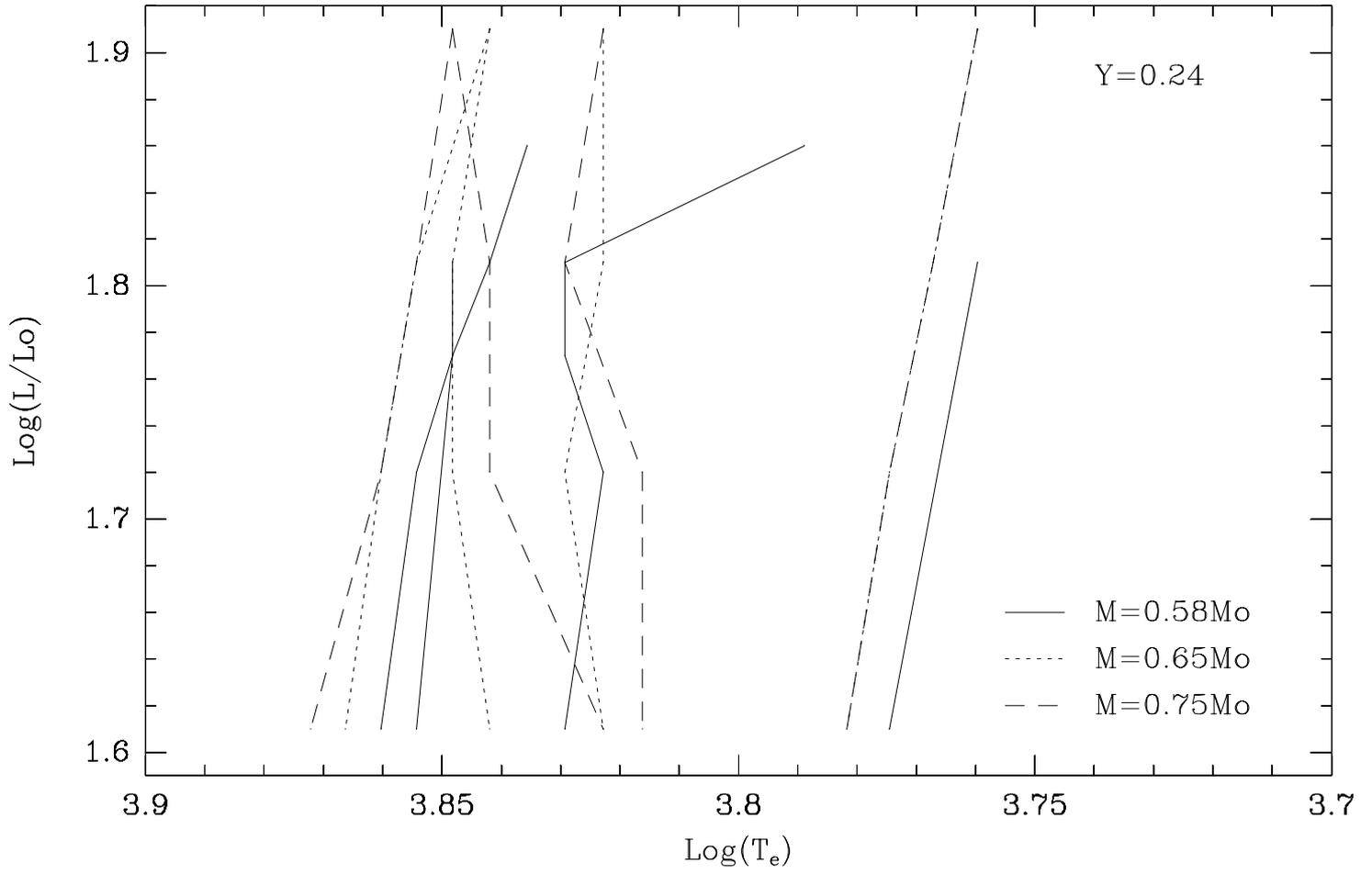

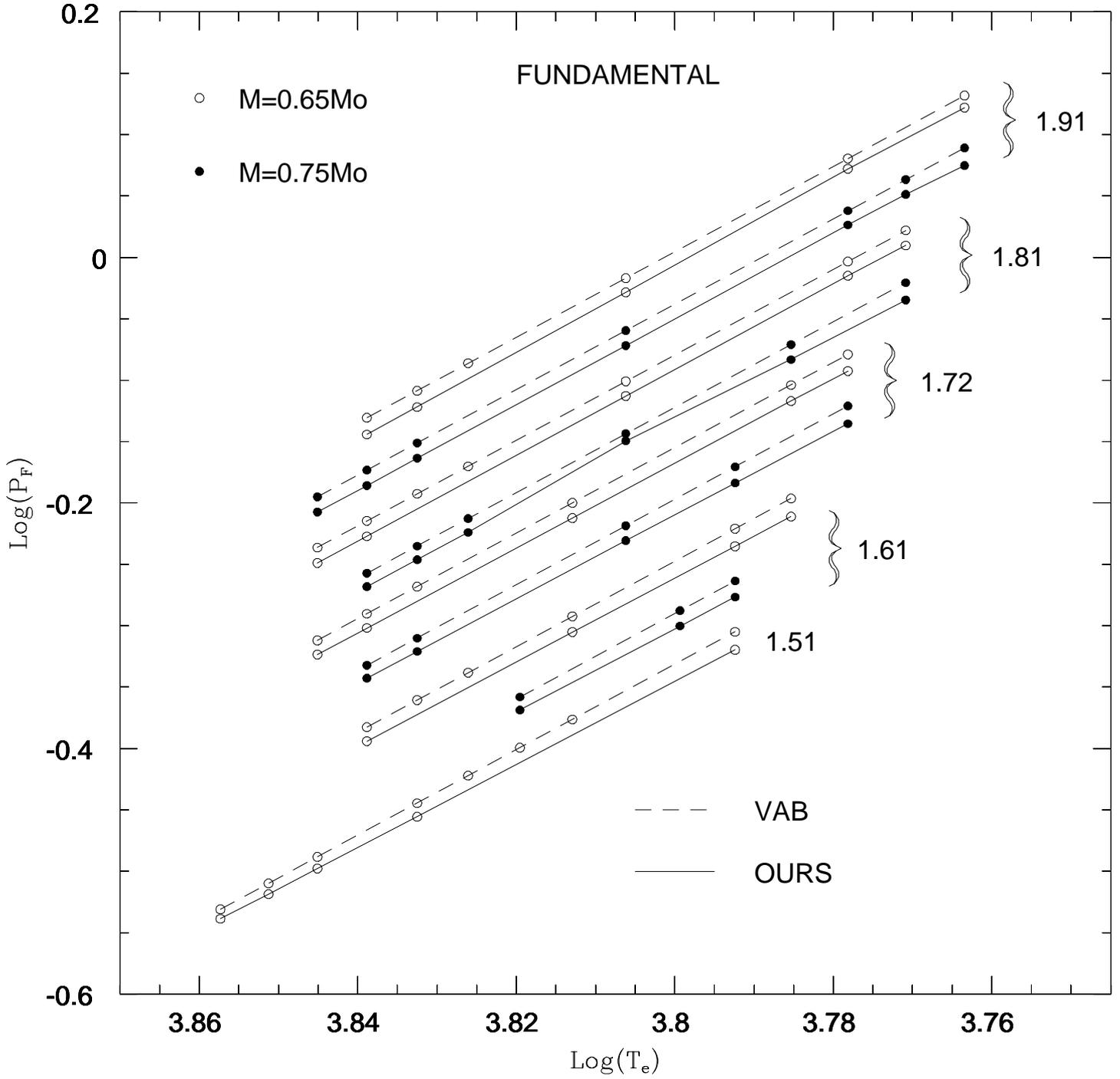

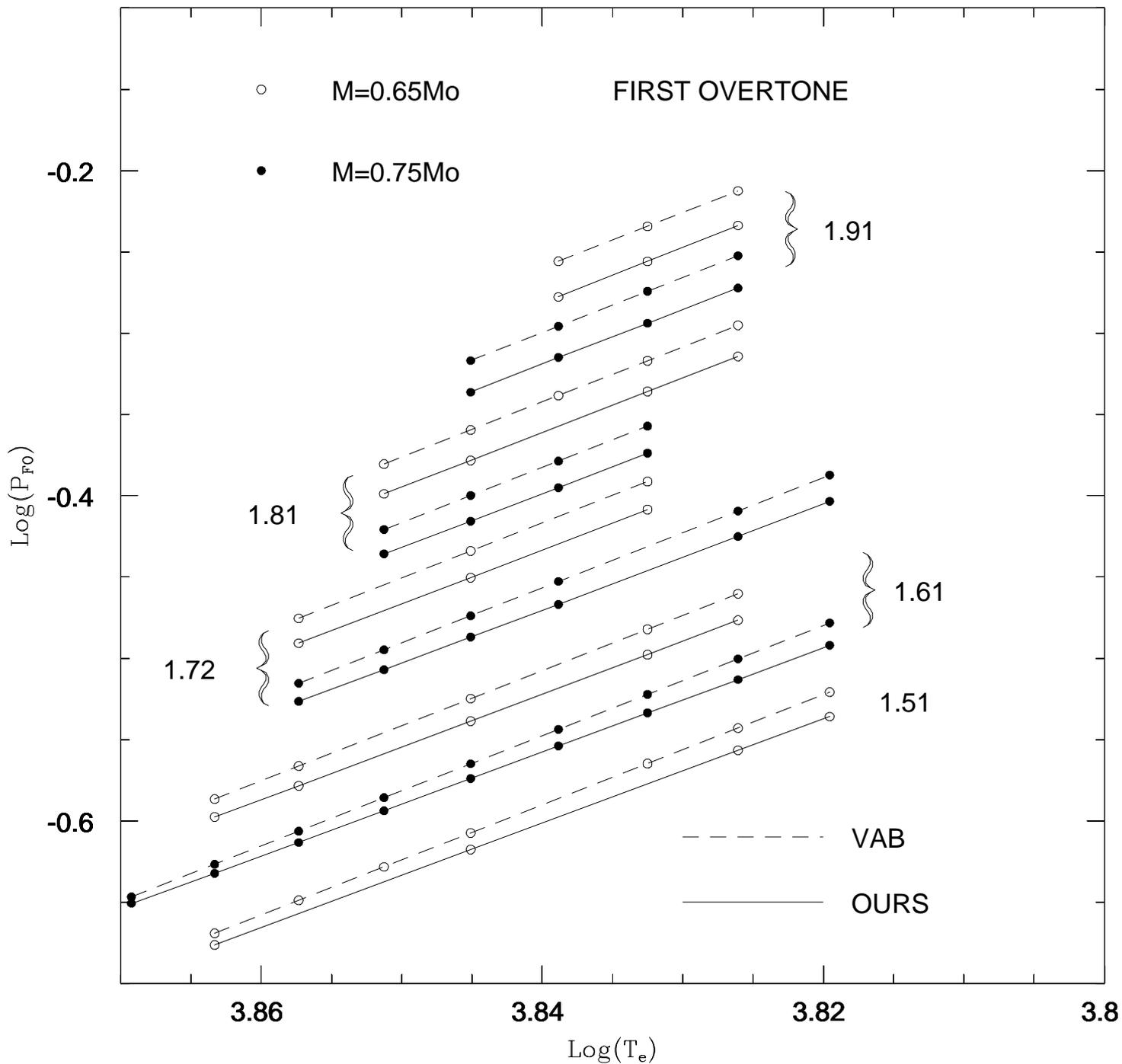

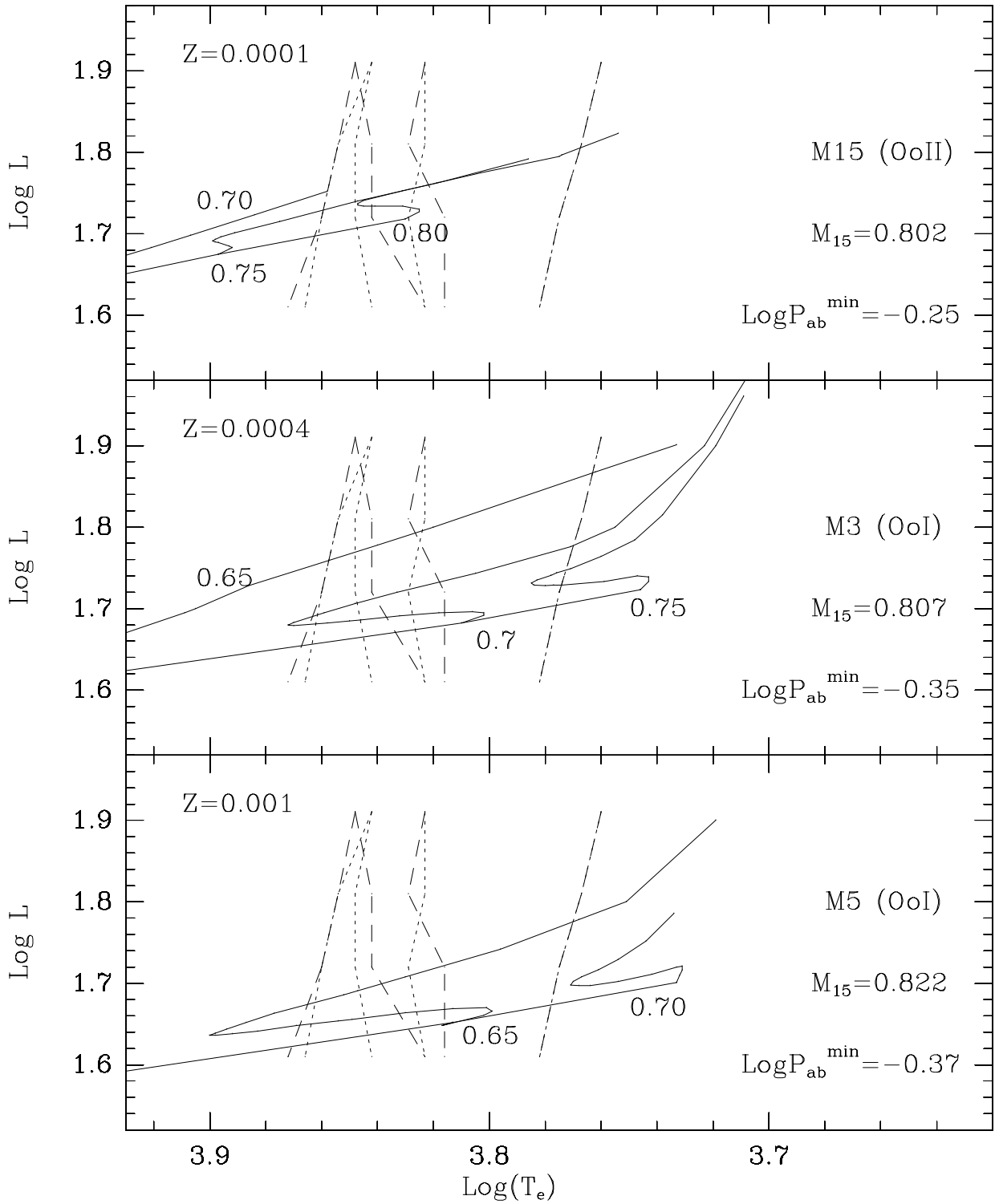

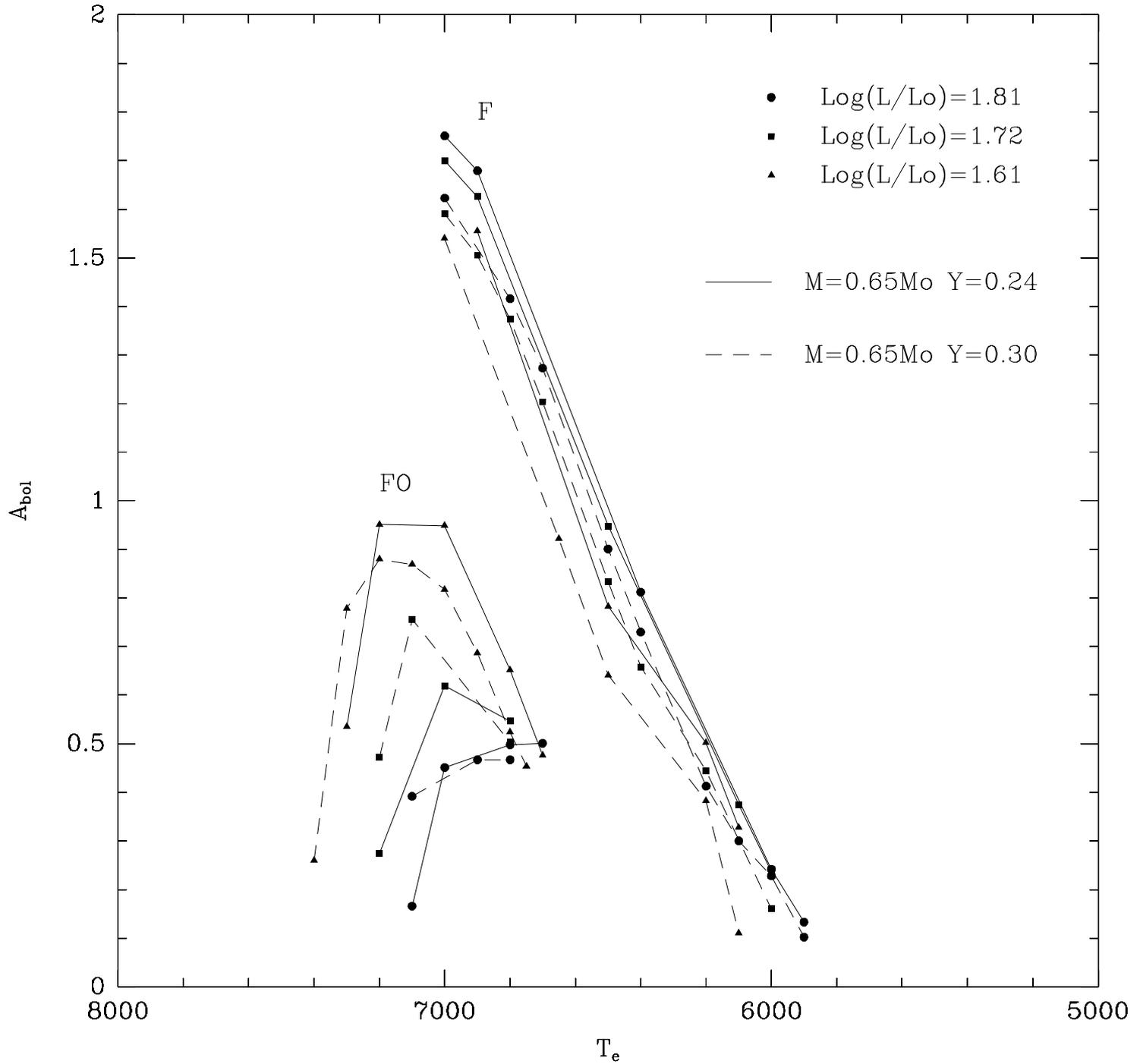

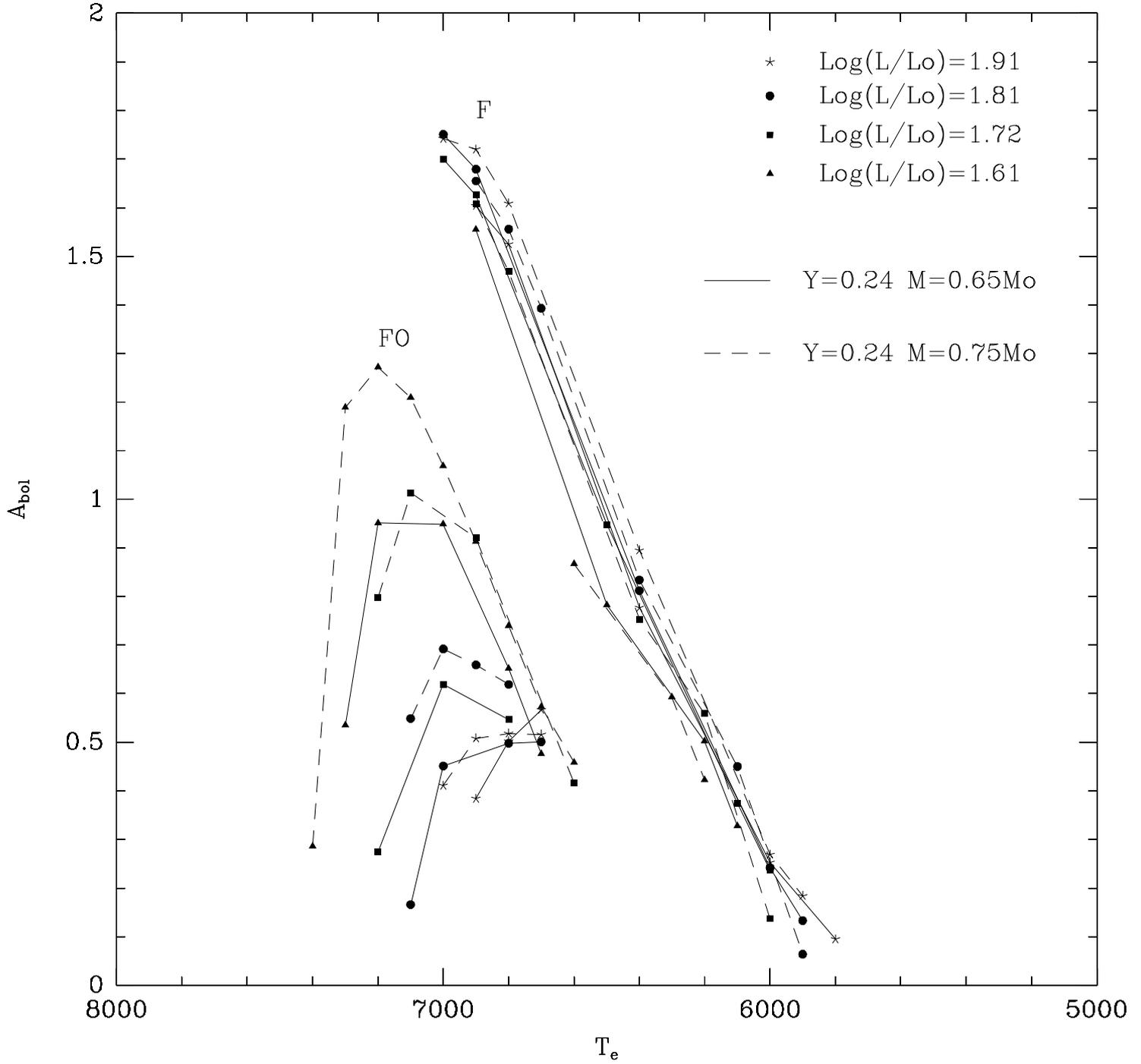

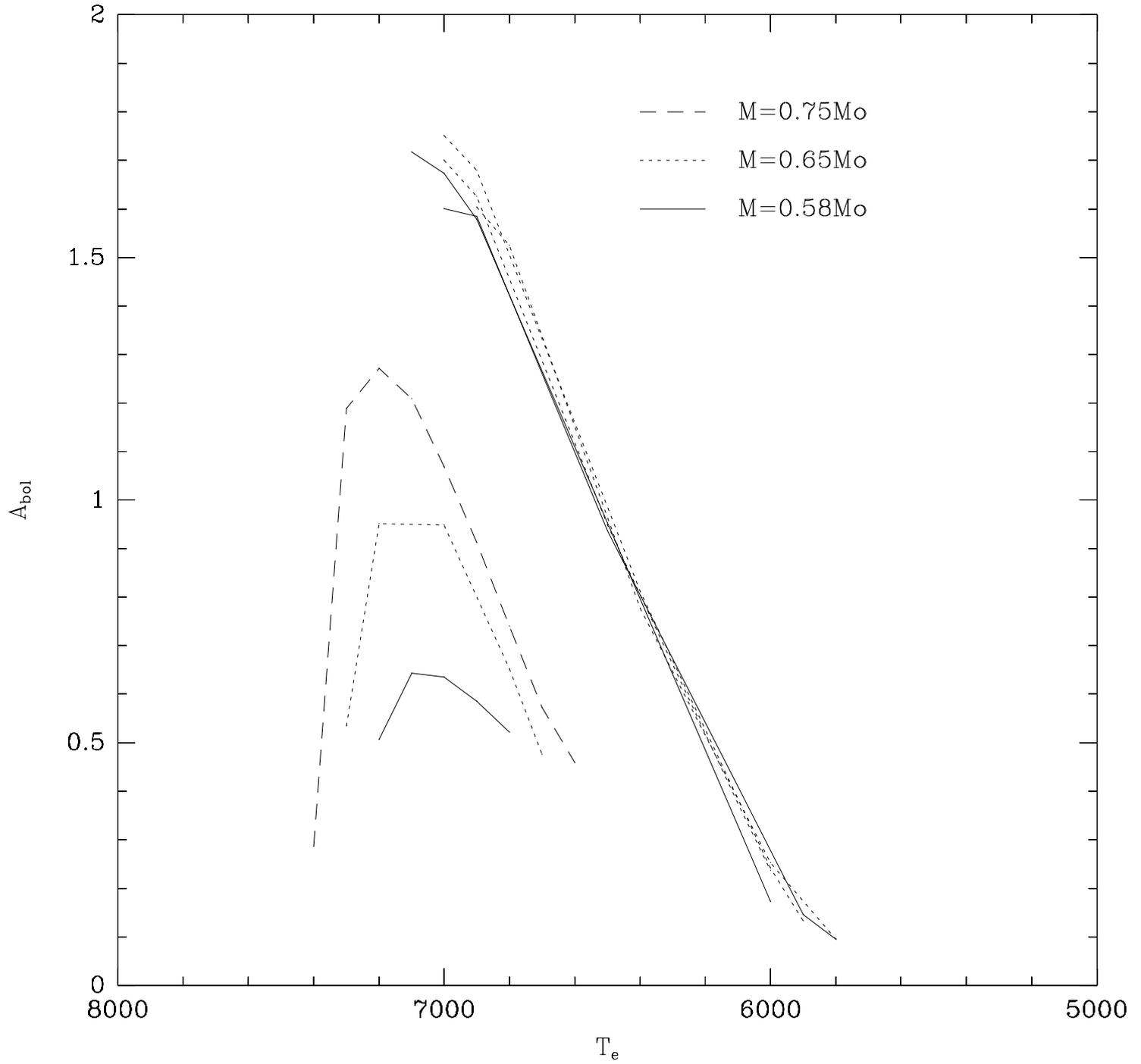

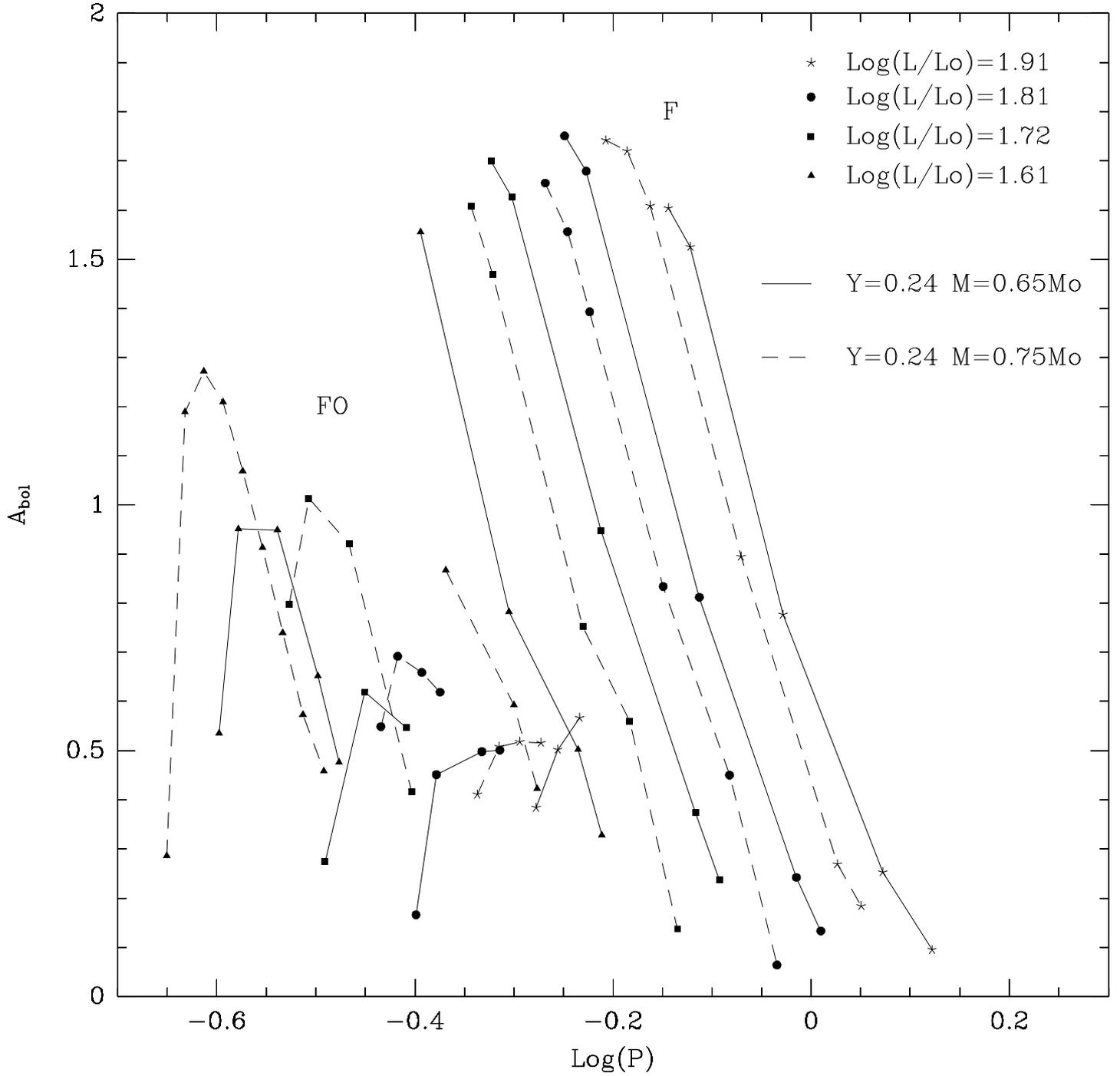

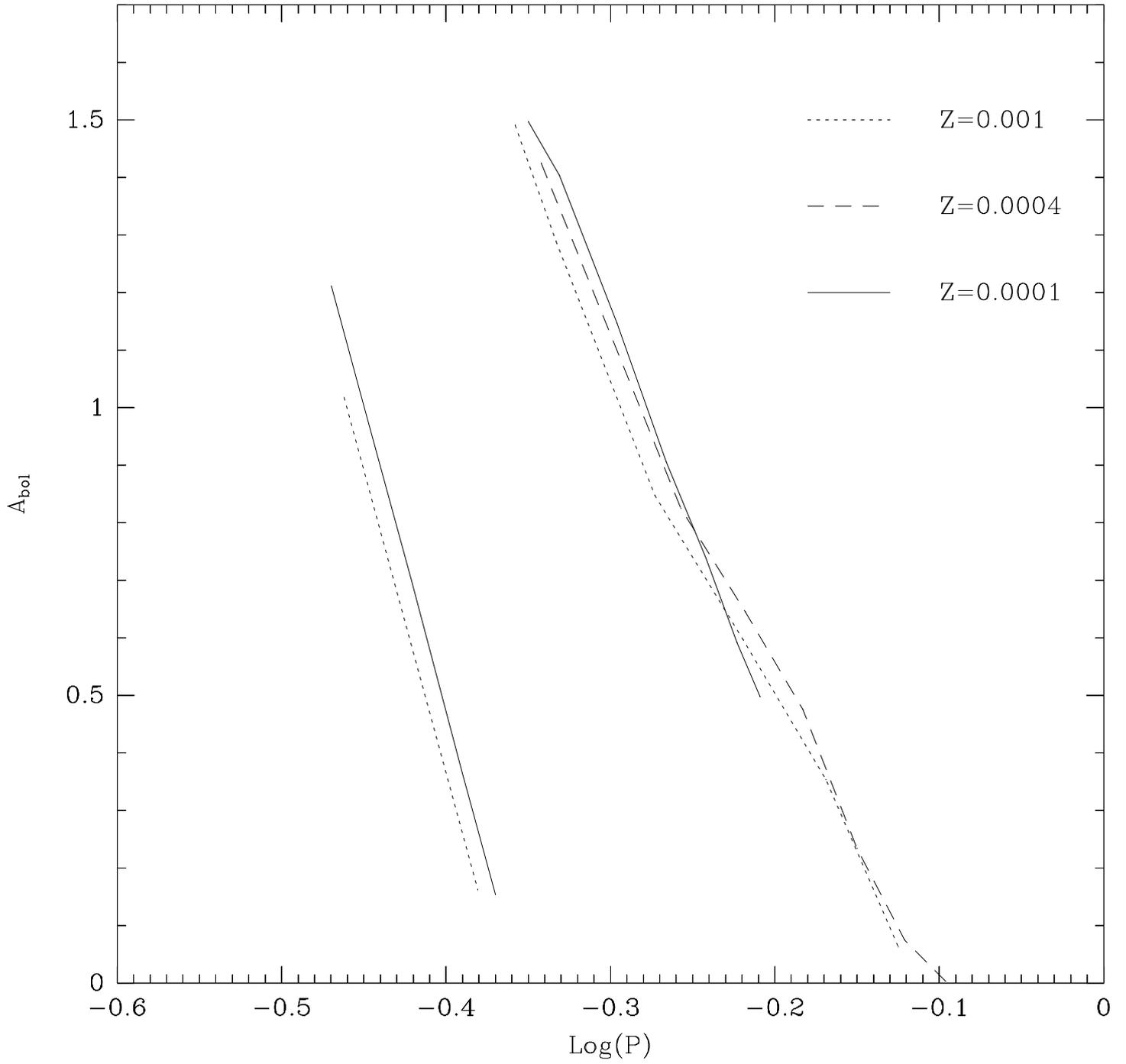

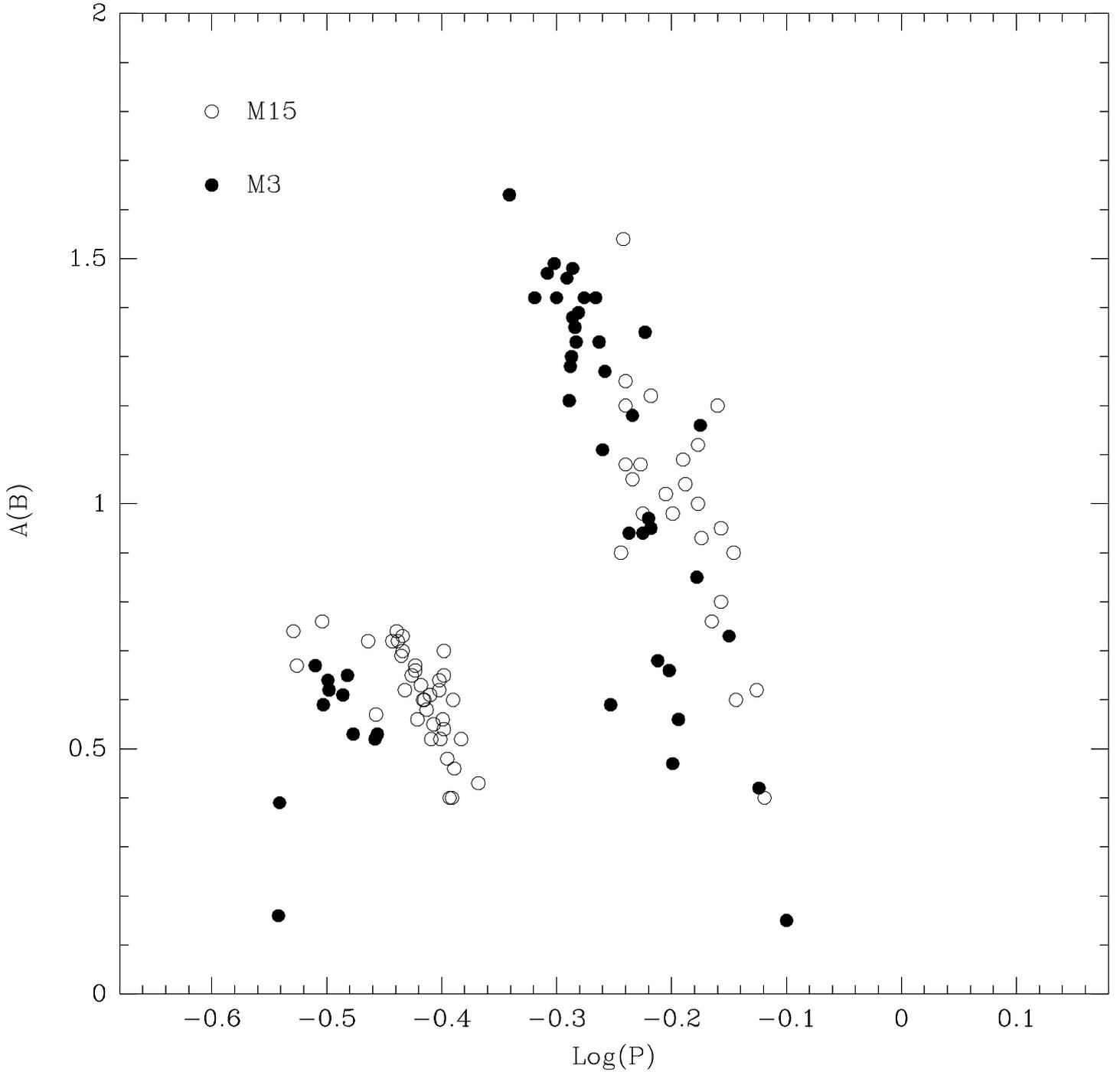

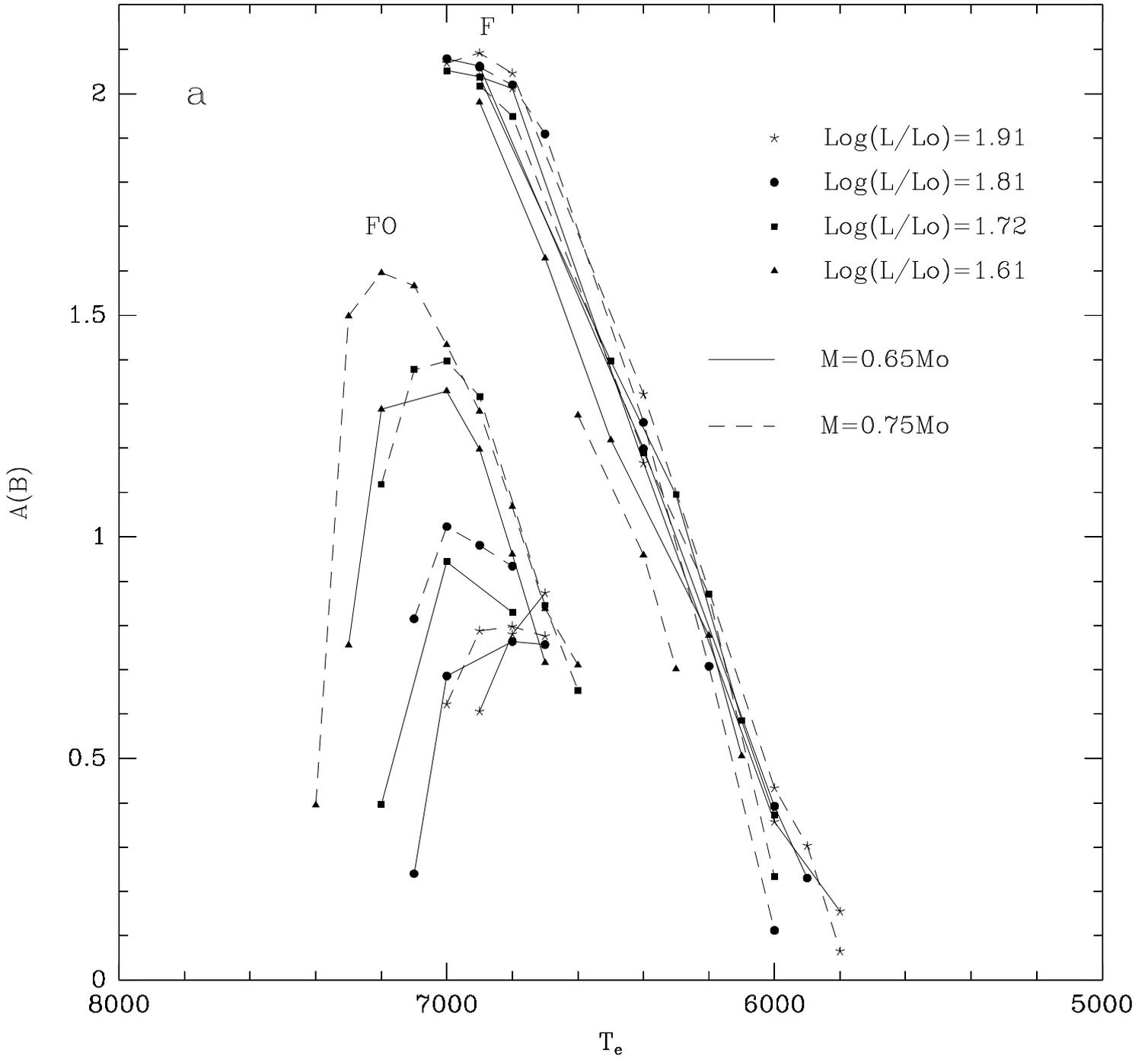

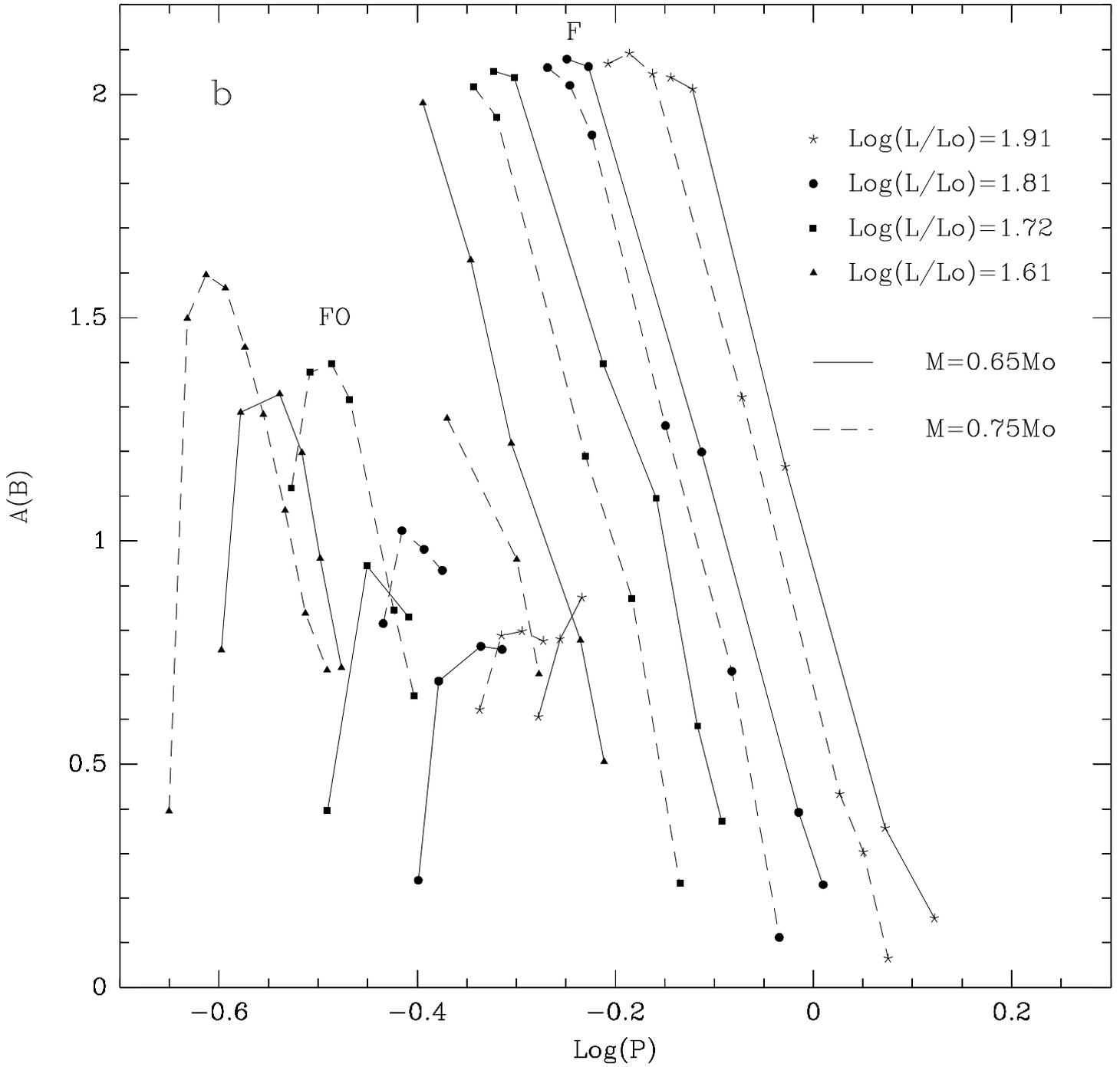

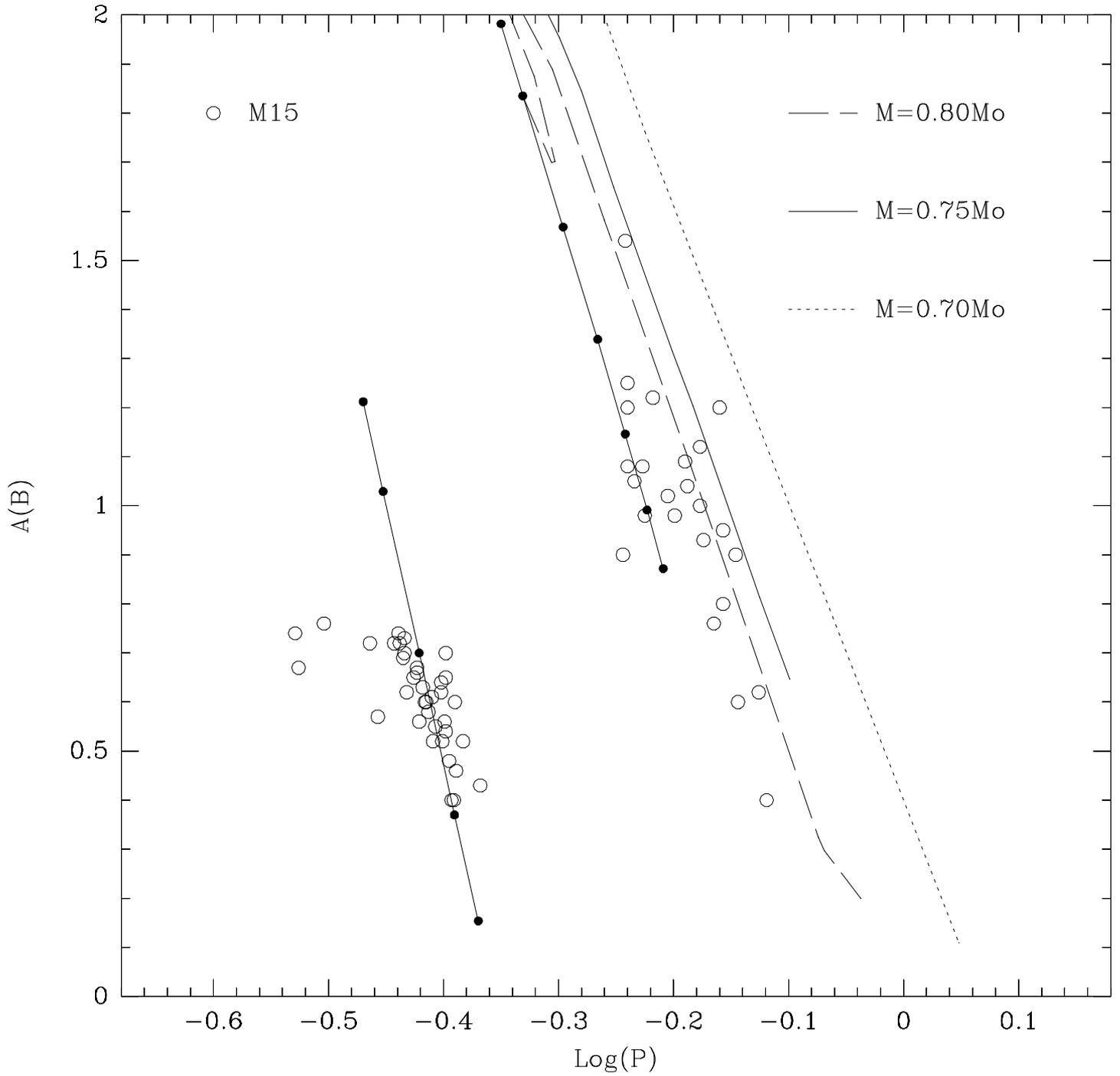

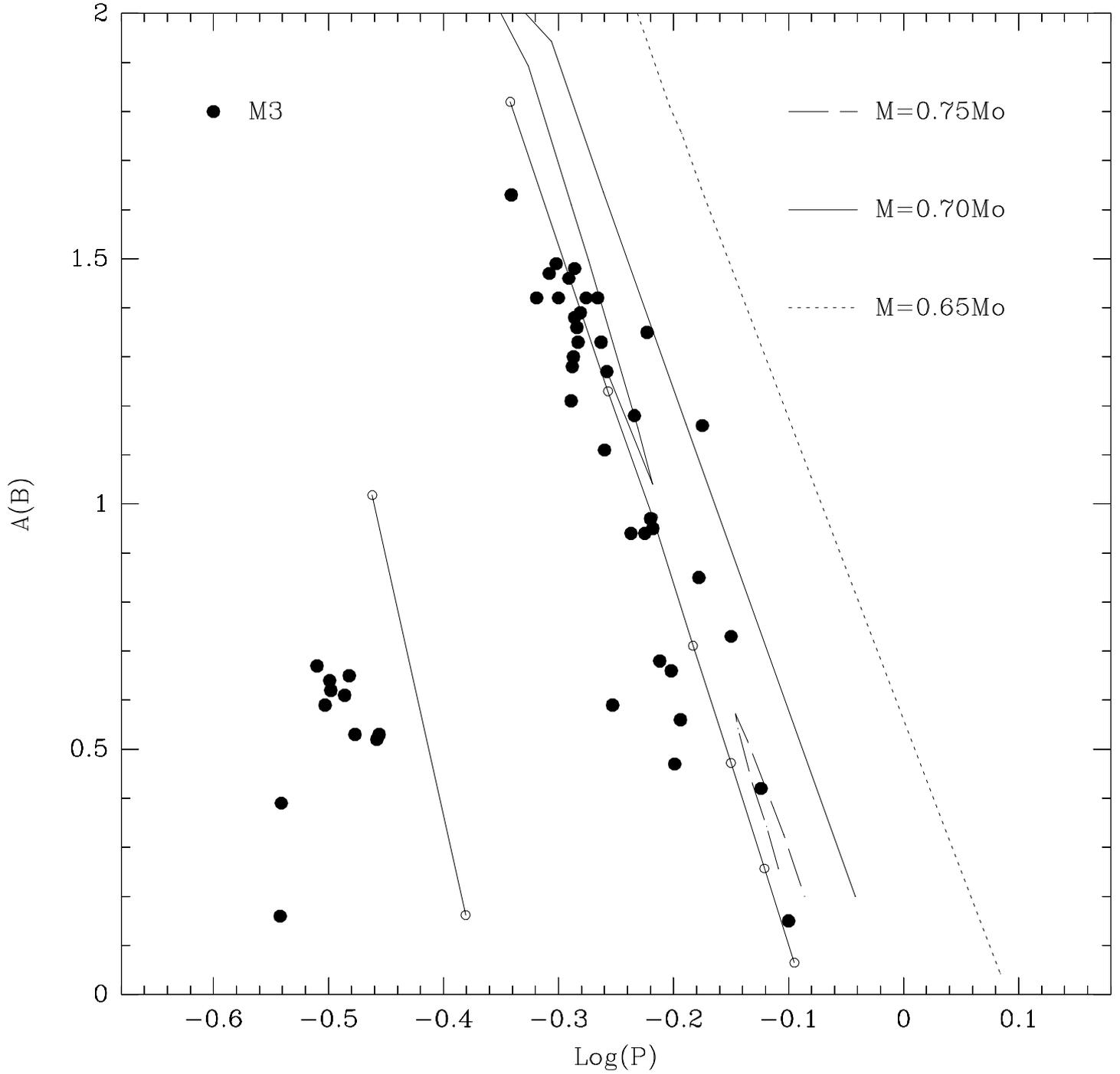

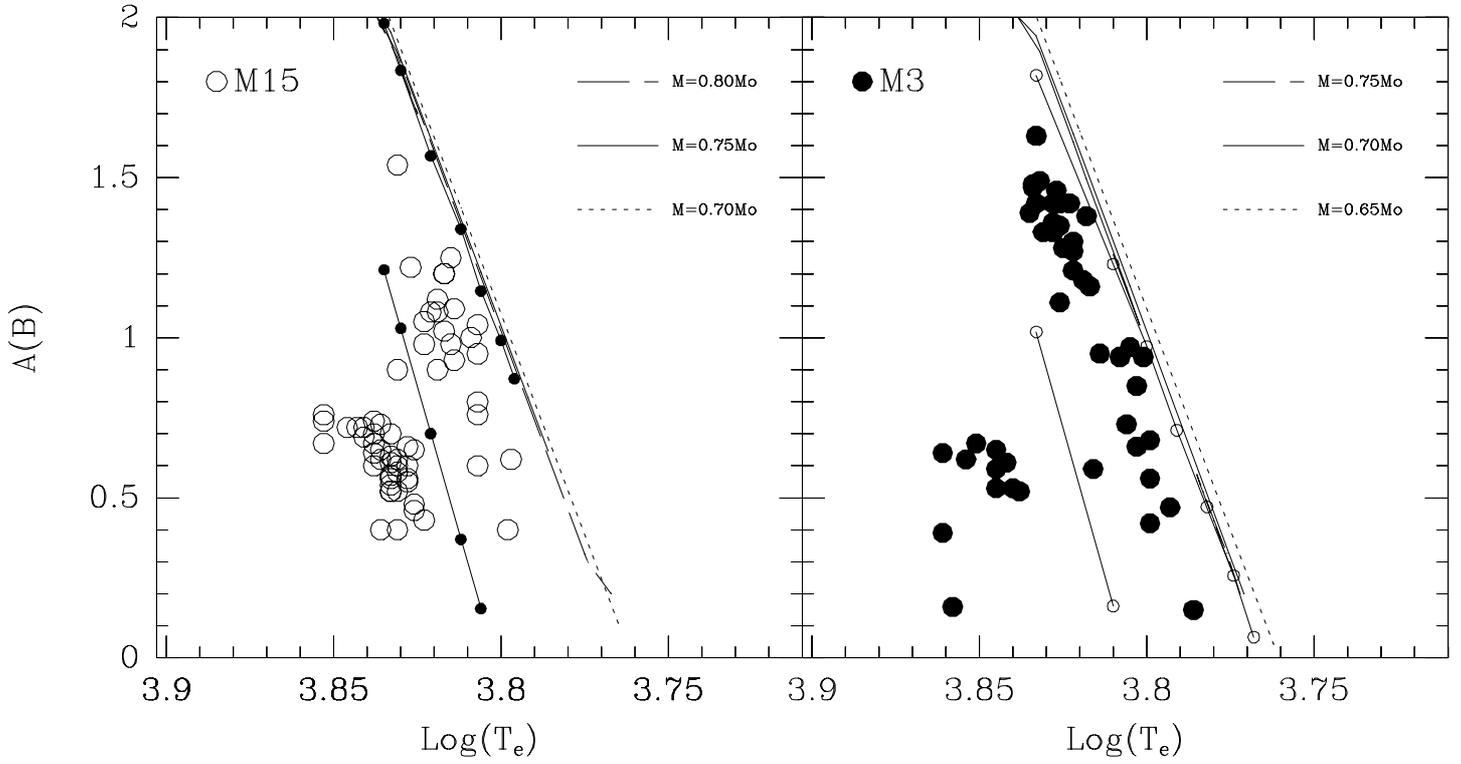